\pdfoutput=1
\documentclass[twocolumn]{aastex701}
\usepackage[T1]{fontenc}
\usepackage{microtype}
\usepackage{CJKutf8}
\newcommand{\cjk}[1]{#1}
\newenvironment{CJKwrap}{
\begin{CJK*}{UTF8}{gbsn}}{
\end{CJK*}}

\usepackage{xspace}
\usepackage{xcolor}
\definecolor{darkgreen}{RGB}{0,100,0}
\usepackage{physics}
\usepackage{siunitx}
\usepackage{ulem}
\usepackage{soul}
\usepackage{etoolbox}

\newcommand{\ocen}{$\omega$\,Cen\xspace}
\newcommand*\chem[1]{\ensuremath{\mathrm{#1}}}
\newcommand{\FeH}{\ensuremath{[\chem{Fe}/\chem{H}]}}

\submitjournal{ApJ}
\shorttitle{Subpopulation Kinematics in Omega Centauri}
\shortauthors{Smith et al.}

\graphicspath{{./figures/}}

\begin{document}

\begin{CJKwrap}

  \title{oMEGACat. XI. Kinematic Characterization of Omega Centauri's Subpopulations: \\  A Flattened Primordial Component }
  \correspondingauthor{Peter J. Smith}
  \author[orcid=0000-0002-7489-5244,sname='Smith']{Peter J. Smith}
  \affiliation{Max Planck Institute for Astronomy, K\"onigstuhl 17, D-69117 Heidelberg, Germany}
  \affiliation{Department of Physics and Astronomy, University of Heidelberg, Im Neuenheimer Feld
  226, D-69120, Heidelberg, Germany}
  \email[show]{pesmith@mpia.de}

  \author[0000-0002-6922-2598]{N. Neumayer}
  \affiliation{Max Planck Institute for Astronomy, K\"onigstuhl 17, D-69117 Heidelberg, Germany}
  \email[]{neumayer@mpia.de}

  \author[0000-0003-0248-5470]{A. C. Seth}
  \affiliation{Department of Physics and Astronomy, University of Utah, Salt Lake City, UT 84112, USA}
  \email[]{aseth@astro.utah.edu}

  \author[0009-0005-8057-0031]{C. Clontz}
  \affiliation{Max Planck Institute for Astronomy, K\"onigstuhl 17, D-69117 Heidelberg, Germany}
  \email[]{clontz@mpia.de}

  \author[0000-0002-5844-4443]{M. H\"aberle}
  \affiliation{European Southern Observatory, Karl-Schwarzschild-Straße 2, 85748 Garching, Germany}
  \email[]{maximilian.haberle@eso.org}

  \author[0000-0001-8052-969X]{S. O. Souza}
  \affiliation{Max Planck Institute for Astronomy, K\"onigstuhl 17, D-69117 Heidelberg, Germany}
  \email[]{s-souza@mpia.de}

  \author[0000-0003-2512-6892]{Z. Wang (\cjk{王梓先})}
  \affiliation{Department of Physics and Astronomy, University of Utah, Salt Lake City, UT 84112, USA}
  \email[]{wang.zixian.astro@gmail.com}

  \author[0000-0003-4546-7731]{G. van de Ven}
  \affiliation{Department of Astrophysics, University of Vienna, T\"urkenschanzstrasse 17, 1180 Wien, Austria}
  \email[]{glenn.vandeven@univie.ac.at}

  \author[0000-0002-2941-4480]{M. S. Nitschai}
  \affiliation{Max Planck Institute for Astronomy, K\"onigstuhl 17, D-69117 Heidelberg, Germany}
  \email[]{nitschai@mpia.de}

  \author[0000-0001-9673-7397]{M. Libralato}
  \affiliation{INAF - Osservatorio Astronomico di Padova, Vicolo dell'Osservatorio 5, Padova I-35122, Italy}
  \email[]{mattia.libralato@inaf.it}

  \author[0000-0002-0160-7221]{A. Feldmeier-Krause}
  \affiliation{Department of Astrophysics, University of Vienna, T\"urkenschanzstrasse 17, 1180 Wien, Austria}
  \email[]{anja.krause@univie.ac.at}

  \author[0000-0001-6604-0505]{S. Kamann}
  \affiliation{Astrophysics Research Institute, Liverpool John Moores University, 146 Brownlow Hill, Liverpool L3 5RF, UK}
  \email[]{s.kamann@liverpool.ac.uk}

  \author[0000-0003-1820-7985]{F. I. Aros}
  \affiliation{Department of Astrophysics, University of Vienna, T\"urkenschanzstrasse 17, 1180 Wien, Austria}
  \email[]{francisco.aros@univie.ac.at}

  \author[0000-0001-6785-9336]{R. Leaman}
  \affiliation{Department of Astrophysics, University of Vienna, T\"urkenschanzstrasse 17, 1180 Wien, Austria}
  \email[]{ryan.leaman@univie.ac.at}

  \author[0000-0003-3858-637X]{A. Bellini}
  \affiliation{Space Telescope Science Institute, 3700 San Martin Dr., Baltimore, MD, 21218, USA}
  \email[]{bellini@stsci.edu}

  \author[0000-0002-1670-0808]{R. Pechetti}
  \affiliation{Nicolaus Copernicus Astronomical Center of the Polish Academy of Sciences, ul. Bartycka 18, 00-716 Warsaw, Poland}
  \email[]{rpechetti@camk.edu.pl}

  \author[0000-0002-1212-2844]{M. Alfaro-Cuello}
  \affiliation{Centro de Investigaci\'on en Ciencias del Espacio y F\'isica Te\'orica, Universidad Central de Chile, La Serena 1710164, Chile}
  \email[]{mayte.alfaro@ucentral.cl}

  \collaboration{all}{(The oMEGACat Collaboration)}

  \begin{abstract}

    Omega Centauri (\ocen; NGC~5139), the likely stripped nucleus of a dwarf galaxy accreted by the
    Milky Way, is one of the few galactic nuclei resolvable into individual stars. Because of its
    long relaxation time, the kinematics of its stars may still retain memory of how this nucleus
    formed. We combine HST proper motions and MUSE line-of-sight velocities from the oMEGACat survey
    to compare populations defined by abundance variations (the chromosome map), metallicity, and
    age within the half-light radius. The chromosome-map division separates two dynamically distinct
    components. The primordial stars have a flattened dispersion field, with kinematic ellipticity
    $\epsilon \approx 0.7$ ($\sigma_{\rm minor} / \sigma_{\rm major} \approx 0.8$), which we recover
    independently in the line-of-sight velocities and the proper motions. The enriched stars are
    consistent with $\epsilon=0$ in the line-of-sight velocities and are no more flattened than the
    full sample in the proper motions, and the two components show no strong difference in rotation.
    Metallicity bins show no comparable structure, because each bin mixes stars from both
    components, and the age--metallicity sample covers only the cluster core, where it cannot test
    for the flattening. The present-day central concentration of the most enriched (P2) stars
    suggests that they formed more centrally concentrated than the primordial stars, independent of
    assembly scenario. These results favor a picture in which the primordial stars retain the
    flattened orbital structure of the progenitor, while the enriched stars fell inward as dense
    clumps or formed in compact, centrally concentrated starbursts.
  \end{abstract}

  \keywords{\uat{Globular star clusters}{656}, \uat{Star clusters}{1567}, \uat{Stellar
    dynamics}{1596}, \uat{Intermediate-mass black holes}{816},
  \uat{Stellar populations}{1622}, \uat{Stellar kinematics}{1608}, \uat{Galaxy nuclei}{609}}

  \section{Introduction} \label{sec:intro}

  Omega Centauri (\ocen; NGC~5139) is the most massive globular cluster in the Milky Way (MW), with
  a mass of approximately $3$--$4 \times 10^6\,M_\sun$ \citep{Baumgardt2018, Dickson2023}. Unlike
  typical globular clusters (GCs), \ocen exhibits complex stellar populations, spanning more than 1
  dex in metallicity \citep{Johnson2010a, Nitschai2023a, Nitschai2024a}, and an age spread of
  $\sim$3~Gyr \citep{Villanova2014,Clontz2024b}. These characteristics strongly suggest that \ocen
  is the remnant nuclear star cluster (NSC) of a dwarf galaxy that was accreted and stripped by the
  MW \citep{Lee1999,Bekki2003,Neumayer2020,Pfeffer2021,Souza2026a}. Furthermore, recent astrometric
  detections of fast-moving central stars support the presence of an intermediate-mass black hole
  (IMBH) of $\gtrsim 8200\,M_\sun$ at its core \citep{Haberle2024}, a hallmark of galactic nuclei.
  Understanding the formation and evolution of \ocen can therefore constrain the processes governing
  NSC formation and the assembly history of the MW (e.g., the Gaia-Enceladus/Sausage (GSE) merger;
    \citealt{Massari2019,Limberg2022}, which \citealt{Laporte2026} date to $\sim$11~Gyr ago and
  associate with a coeval burst of GC formation).

  Decades of photometric and spectroscopic studies have revealed that \ocen hosts multiple, distinct
  populations of stars \citep[e.g.,][]{Norris1995, Lee1999, Bedin2004, Johnson2010a, Marino2011,
  Bellini2017a}. The oMEGACat survey\footnote{\url{https://omegacatalog.github.io/}} \citep{Nitschai2023a, Haberle2024b} has since mapped the
  metallicity \citep{Nitschai2024a}, age \citep{Clontz2024b}, helium enhancement
  \citep{Clontz2025a}, and chemical abundance \citep{Wang2026a} distributions of these populations,
  as well as the intracluster medium \citep{Wang2025b}.

  Using the pseudo-color ``chromosome map'' (ChM) \citep{Milone2015, Milone2017} constructed from the
  combination of the F275W, F336W, F435W and F814W Hubble Space Telescope (HST) filters, the cluster
  separates into three primary groups: the primordial (P1) stars, and the chemically enriched
  intermediate (Im) and P2 stars \citep{Clontz2026, Dondoglio2026}, where we follow the naming of
  \citet{Clontz2026}. The P2 population is characterized by extreme helium enhancement ($\Delta Y
  \sim 0.15$), nitrogen and sodium enrichment, and oxygen depletion \citep{Norris2004, Piotto2005,
  Johnson2010a, Marino2011, Marino2012, Clontz2025a, Wang2026a}. By linking the chromosome map on
  the red giant branch (RGB) to the subgiant branch (SGB) age--metallicity relation (AMR),
  \citet{Clontz2026} demonstrated a $\sim$1~Gyr age gap between the P1 populations and the P2
  populations across all metallicities, which requires an extended formation history
  \citep[e.g.,][]{Romano2007, Romano2010, Marino2012, Bellini2017a}. The true population count is
  likely even finer than this three-way split: \citet{Bellini2017a} identify at least 15
  subpopulations along the main sequence (MS), \citet{Akbaba2026a} recover ten distinct populations
  from clustering in abundance space, while \citet{Clontz2026} identify 14 subpopulations through
  data-driven clustering in photometry and metallicity.

  The labels P1 and P2 echo the primordial and enriched populations of typical GCs,
  but differ from them in important ways: P2's helium enhancement lies at the extreme of the range
  seen in GCs, and both P1 and P2 span a wide range in \FeH\ \citep{Johnson2010a,
  Marino2011} and in age \citep{Clontz2026}, whereas the populations of typical clusters share a
  single metallicity and are close to coeval \citep{Bastian2018}. The scenarios proposed for the
  multiple populations of typical clusters therefore may not apply to \ocen's P1 and P2.
  
  Stellar kinematics probe the assembly history of stellar systems. In particular, the multiple stellar
  populations observed in typical GCs have long been known to exhibit distinct spatial
  distributions \citep[e.g.,][]{Lardo2011, Leitinger2023a, Mehta2025} and kinematic properties
  \citep[e.g.,][]{Cordero2017, Milone2018b, Libralato2023a, Dalessandro2024a, Cordoni2025a,
  Leitinger2025}. These differences can provide clues about the formation mechanisms of the various
  populations, such as whether they formed in situ or were accreted from external sources (e.g.,
    pristine gas from the surroundings, \citealt{DErcole2008}; or infalling star clusters,
  \citealt{Antonini2012}).

  Standard models predict that P2 forms from enriched ejecta left behind by P1 stars, which settle
  into a compact, rotating, centrally concentrated configuration before undergoing star formation
  \citep{Bekki2010, Mastrobuono-Battisti2013}. Hydrodynamical simulations of second-generation
  formation from asymptotic giant branch (AGB) ejecta mixed with accreted low-density pristine gas
  likewise produce a compact, centrally concentrated second generation~\citep{Yaghoobi2022}. The
  specific enrichment channel is debated, with proposed sources including AGB
  stars~\citep[e.g.,][]{DErcole2008, DAntona2026}, interacting binaries~\citep{DeMink2009},
  fast-rotating massive stars~\citep{Decressin2007}, and supermassive or accreting extremely massive
  stars~\citep{Denissenkov2014, Gieles2018a, Gieles2025a}~\citep[see][for critical
  assessments]{Renzini2015, Bastian2018}, and most predict that P2 is more centrally concentrated
  than P1, since enriched ejecta settle toward the cluster center. Recent observations, however,
  present a more mixed view: among dynamically young clusters, some have a centrally concentrated
  enriched population, some a centrally concentrated primordial population, and some fully mixed
  populations \citep{Leitinger2023a}.

  Large proper-motion and MUSE samples now show that kinematic differences between the populations
  are common. In most clusters the P2 stars have enhanced radial anisotropy compared to the P1
  stars, which is consistent with the P2 stars being more centrally concentrated \citep{Milone2018b,
  Libralato2023a, Dalessandro2024a, Cordoni2025a}. The evidence for a rotation difference is
  similarly mixed, with no universal, consistent difference in rotation between P1 and P2 stars
  across clusters \citep{Cordero2017, Dalessandro2024a, Leitinger2025}. These differences are weaker
  in dynamically older clusters \citep{Dalessandro2024a, Cordoni2025a}, and some clusters show no
  significant difference or the opposite sense \citep[e.g.,][]{Martens2023a}.

  The standard formation models were developed in the context of GC formation, not NSCs embedded
  within a dwarf galaxy potential, so their applicability to \ocen is not guaranteed. Nuclear star
  clusters commonly rotate in the same sense as their host galaxies, and some are significantly more
  flattened and rapidly rotating than GCs \citep{Neumayer2020}. The host galaxy likely imposes
  large-scale angular momentum, meaning P1 itself may have formed in a flattened configuration,
  unlike the isotropic primordial stars assumed in standard multiple population formation models for
  GCs. Crucially, different formation pathways are expected to leave distinct kinematic imprints,
  including differences in population-dependent spatial flattening, that can survive the long-term
  dynamical evolution of old clusters and thereby allow different scenarios to be distinguished
  observationally \citep{Mastrobuono-Battisti2013, Henault-Brunet2015, Mastrobuono-Battisti2016,
  Cordero2017} \citep[see also][]{Tiongco2019, Vesperini2021, Aros2025a}. In \ocen, each episode of
  star formation or accretion may likewise have left a residual kinematic signature.

  \ocen is particularly well suited to this search, because of its long relaxation time (the
    half-mass relaxation time is $21$~Gyr; \citealt{Baumgardt2018, Baumgardt2021}, updated online
  catalog),\footnote{\url{https://people.smp.uq.edu.au/HolgerBaumgardt/globular/}} which allows
  it to retain memory of its initial conditions for longer periods compared to typical globular
  clusters \citep{Mastrobuono-Battisti2013}. This makes it possible to observe kinematic
  differences between populations that may have formed under different conditions or at different
  times.

  Previous studies of \ocen have revealed complex kinematic structures, including a kinematically
  cold, disk-like component at 1--3\arcmin~\citep{VanDeVen2006}, and a counter-rotating core
  \citep{Pechetti2024a}. Beyond the half-light radius ($r_h$), population-resolved studies find that the
  primordial stars rotate faster than the enriched stars at $\sim3.5\,r_h$ \citep{Bellini2018}, and
  have a broader angular-momentum distribution and weaker radial anisotropy at $0.9$--$2.3\,r_h$
  \citep{Ziliotto2026}. Out to large radii, aluminum (Al)-poor stars are also closer to isotropy than Al-rich
  stars, although the two groups rotate alike and do not map one-to-one onto P1 and P2
  \citep{Pagnini2026a}. Recently, \citet{Haberle2025} provided a comprehensive 3D kinematic analysis
  of the inner half-light radius of the cluster using the oMEGACat dataset, investigating the
  rotation profile, velocity anisotropy, and degree of energy equipartition in the cluster.
  \citet{Haberle2026a} extended the line-of-sight (LOS) rotation curve to $\sim3\,r_h$ with
  integrated-light spectroscopy from the SDSS-V Local Volume Mapper \citep{Drory2024a,
  Kollmeier2026}. While \citet{Haberle2025} looked for metallicity-dependent kinematics, they found
  no significant differences when splitting the sample into \FeH\ bins, motivating the more detailed
  population definitions we use here.
  
  In this paper, we compare the kinematics of the multiple populations of \ocen\ within the
  half-light radius, using high-precision HST proper motions (PMs; \citealt{Haberle2024b}) and MUSE
  LOS velocities \citep{Nitschai2023a} from the oMEGACat survey. We divide the stars by
  chromosome-map position, by \FeH, and by position on the AMR, and model the rotation, dispersion,
  and anisotropy of each population in 1D and 2D. Section~\ref{sec:data} describes the data and
  quality cuts, Section~\ref{sec:methodology} the population definitions and kinematic models,
  Section~\ref{sec:results} the results, Section~\ref{sec:discussion} their implications for the
  formation of \ocen, and Section~\ref{sec:conclusions} our conclusions.

  \section{Data} \label{sec:data}

  The basis of this work is the oMEGACat catalog, a
  joint HST+MUSE survey of \ocen. The catalog combines HST photometry and proper motions from
  \citet{Haberle2024b} with MUSE spectroscopy from \citet{Nitschai2023a}, and each dataset covers
  the central half-light radius of the cluster ($r_h \approx 4.8\arcmin\approx7.6$~pc;
  \citealt{Baumgardt2018, Baumgardt2021}, updated online catalog).

  \subsection{MUSE}
  The MUSE dataset consists of integral field spectroscopy covering the half-light radius of \ocen.
  The MUSE data are processed with the \texttt{PampelMuse} software \citep{Kamann2013}, which
  extracts individual stellar spectra from the crowded field. Line-of-sight velocities and
  metallicities are measured from these spectra using the \texttt{spexxy} software
  \citep{Husser2016}. The final catalog contains spectra for over 300,000 stars. Typical
  LOS velocity uncertainties for bright RGB stars are $\lesssim 3$~km~s$^{-1}$
  \citep{Nitschai2023a}.

  \subsection{HST}
  The HST dataset provides high-precision proper motions and seven-band photometry (ranging from
  ultraviolet (UV) to near-infrared) within the half-light radius of \ocen. The proper motions of
  1.4 million stars are derived from multi-epoch HST imaging with a typical baseline of 20.1~yr. The
  proper motions are relative to the local mean motion of the cluster stars, which removes any
  plane-of-sky rotation, and achieve a median 1D precision of $\sim 11~\mu$as~yr$^{-1}$ for $m_{\rm
  F625W} \approx 18$ mag, and down to $6.6~\mu$as~yr$^{-1}$ in the highly sampled cluster core
  \citep{Haberle2024b}.

  \subsection{Ages and Chromosome Map}
  We additionally use stellar ages from \citet{Clontz2024b}, who derived precise ages for SGB stars
  in \ocen by fitting the HST photometry with isochrones that adopt a specific C+N+O--\FeH\
  relation. The ages span a range mostly from 10 to 13~Gyr, with median uncertainties of $0.68$~Gyr.
  We also use the chromosome map from \citet{Nitschai2024a}, described in Section~\ref{sec:chromo}.

  \begin{figure*}
    \centering
    \includegraphics[width=\linewidth]{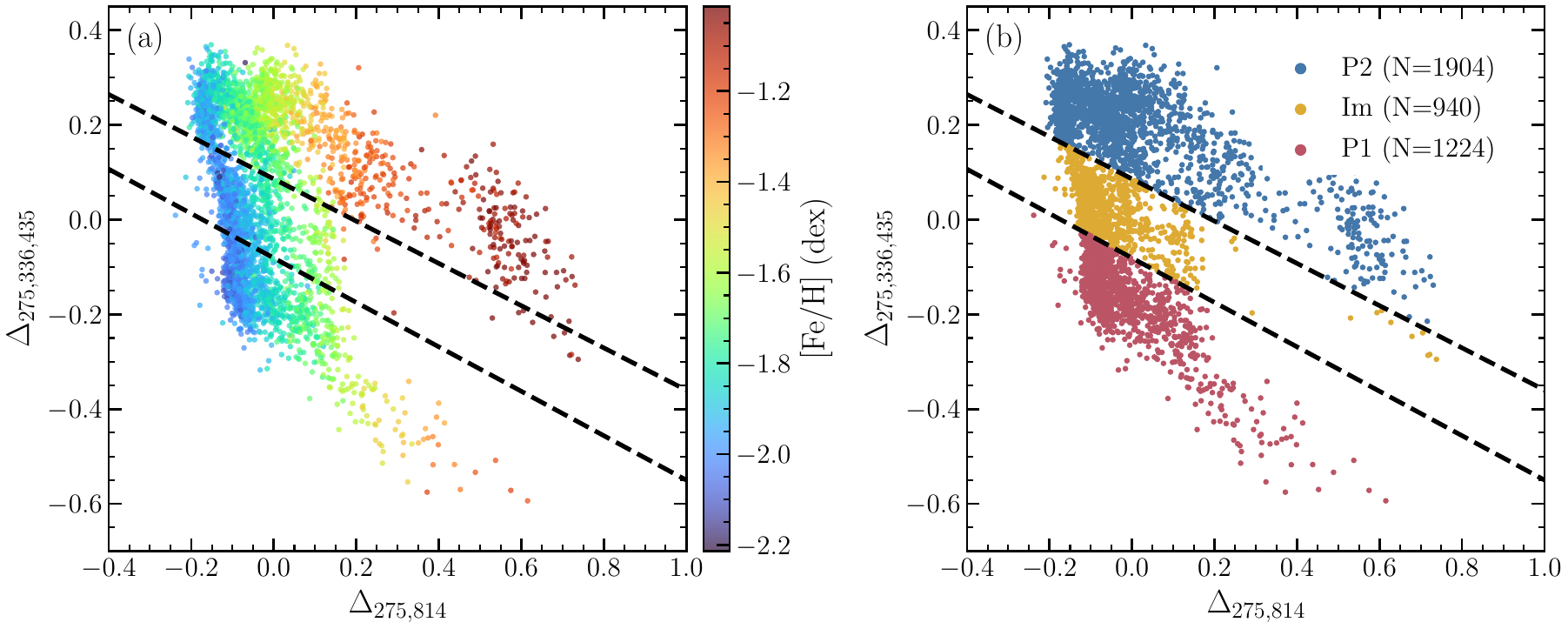}
    \caption{Chromosome map (ChM) of \ocen RGB stars, constructed from the pseudo-color indices
      $\Delta_{275,814}$ and {$\Delta_{275,336,435}$} from HST photometry \citep{Haberle2024b}.
      $\Delta_{275,814}$ is the F275W$-$F814W color and $\Delta_{275,336,435}$ the pseudo-color
      $C_{\rm F275W,F336W,F435W} = ({\rm F275W}-{\rm F336W})-({\rm F336W}-{\rm F435W})$, each
      verticalized relative to fiducial lines bounding the RGB. (a) Stars are color-coded by MUSE
      metallicity \FeH\ \citep{Nitschai2023a}. (b) The three populations used in this work, P1
      (red), Im (yellow), and P2 (blue), are defined: primordial (P1, lower sequence), intermediate
      (Im, middle), and enriched (P2, upper sequence), following \citet{Nitschai2024a} and
      \citet{Clontz2025a}; dashed lines mark the P1/Im and Im/P2 boundaries. The vertical separation
      primarily traces nitrogen enhancement, while the horizontal axis is sensitive to temperature,
      meaning it probes largely metallicity (and helium enhancement)~\citep{Milone2015,
    Clontz2025a}.}
    \label{fig:chromo-def}
  \end{figure*}

  \subsection{Kinematic Quality Cuts}
  \label{sec:kinquality}
  To ensure reliable kinematic measurements, we apply strict quality cuts to the datasets, adopting
  the recommendations from \citet{Haberle2025} for both the HST proper motions and MUSE LOS
  velocities. We select stars with high-quality measurements, removing those with large
  uncertainties or potential contamination from neighboring sources. We apply their catalog flags
  \texttt{selection\_hq\_astrometry\_and\_membership} (HST proper motions and cluster membership)
  and \texttt{selection\_hq\_los} (MUSE LOS velocities). Every star therefore appears in both
  catalogs, so the single membership selection applies to both datasets. Table~\ref{tab:cuts}
  (Appendix~\ref{app:cuts}) gives the number of stars in each subpopulation after each cut, and
  shows that no individual subpopulation is preferentially removed by the quality cuts.

  At this point, the quality cuts are purely kinematic; we apply additional cuts in the following
  sections to ensure that the stars have the necessary chemical abundance and age information to be
  classified into the various subpopulations, each within the magnitude range in which the
  population is defined (Section~\ref{sec:pop_defs}).

  \section{Methodology} \label{sec:methodology}

  To characterize the kinematic properties of the multiple stellar populations in \ocen, we first
  define subpopulations using three complementary methods based on photometric abundance variations,
  \FeH, and age. We then model the kinematics of each subpopulation using 2D spatial maps to infer
  their 3D kinematic structure. In the main text we focus on the 2D fields; the corresponding 1D
  radial profiles are given in Appendices~\ref{app:chromo-1d}--\ref{app:amr}.

  \subsection{Population Definitions} \label{sec:pop_defs}

  We employ three complementary population divisions to test which aspect of \ocen's chemistry
  traces its dynamics: the chromosome map (star-by-star abundance variations), groupings on the AMR,
  and pure \FeH\ quintiles. By comparing the kinematics of these populations, we can constrain the
  formation history of the cluster and the processes that shaped its multiple stellar populations.

  \subsubsection{Chromosome Map (Abundance Variations)} \label{sec:chromo}

  The chromosome map, introduced by \citet{Milone2015}, separates stellar populations based on
  multi-band photometry sensitive to light-element abundance variations. It is built from two
  pseudo-colors, $\Delta_{275,814}$ from $m_{\rm F275W}-m_{\rm F814W}$ and $\Delta_{275,336,435}$
  from $(m_{\rm F275W}-m_{\rm F336W})-(m_{\rm F336W}-m_{\rm F435W})$, each measured relative to
  fiducial lines along the RGB \citep{Piotto2015, Milone2015, Nitschai2024a}. In particular, the
  index $\Delta_{275,336,435}$ is sensitive to carbon, nitrogen, and oxygen abundances through
  molecular bands (OH, NH, CN, and CH features), while $\Delta_{275,814}$ is primarily
  temperature-sensitive and traces helium enrichment indirectly through its effect on stellar
  temperature, while also tracing iron abundance, or total metallicity.

  We adopt the chromosome map from \citet{Nitschai2024a}, with the population boundaries from
  \citet{Clontz2025a}, to define three distinct populations within \ocen: primordial (P1) stars,
  which occupy the lower portion of the map and are characterized by primordial abundance patterns;
  enriched (P2) stars, which occupy the upper portion and exhibit enhanced nitrogen and helium; and
  an intermediate (Im) population that lies between P1 and P2. In Figure~\ref{fig:chromo-def}, we
  illustrate the definition of these populations on the chromosome map, showing the distinct regions
  occupied by P1, P2, and Im stars as well as the range of metallicities covered by each group. In
  addition to C, N, O, and He, the sequences also differ in Al and in heavy elements such as La
  \citep{Johnson2010a, Marino2011, Mason2026, Wang2026a}. Full abundance and age characterizations
  of these groups are given in \citet{Clontz2025a,Wang2026a,Clontz2026}. This separation allows us
  to investigate the kinematic properties of these chemically distinct populations, without
  requiring detailed spectroscopic abundances for every star.

  Beyond the shared kinematic quality cuts (Section~\ref{sec:kinquality}), stars are required to
  have a valid chromosome-map pseudo-color, and to additionally fall on the RGB within $12.9 <
  m_{\rm F814W} < 16.1$, following the RGB selection of \citet{Clontz2026}; the faint limit keeps
  the sample away from the SGB. The final sample has $N=4068$ stars: $N_{\rm P1}=1224$, $N_{\rm
  Im}=940$, and $N_{\rm P2}=1904$.

  \subsubsection{Age--Metallicity Relation (AMR)}
  \label{sec:amr_def}
  We use the AMR measured by \citet{Clontz2024b} to define subpopulations based on stellar ages and
  iron abundances. The AMR traces the chemical evolution of \ocen in time, allowing us to separate
  stars into groups that reflect different formation epochs and enrichment histories. We show in
  Figure~\ref{fig:AMR-def} the distribution of stars in the AMR space. In panel (a), we show a 2D
  histogram, colored by the log-density of points in the diagram, showing the distinct sequences
  present. In panel (b) we define three populations based on their position in the AMR space:
  ``Tight'' stars (in green), which, at fixed metallicity, are the youngest and occupy a narrow
  linear sequence; ``Diffuse'' stars (in orange), which are older at fixed metallicity and exhibit a
  broader spread; and ``Metal-Rich'' stars (in gray), which occupy the most metal-rich portion of
  the AMR and are generally younger. This classification allows us to investigate how age and iron
  abundance correlate with kinematic properties. AMR membership additionally requires a non-null
  fiducial age and \FeH\ from \citet{Clontz2024b}, restricting this sample to SGB stars with
  well-measured ages. After the quality cuts, the samples contain 2573 Tight, 2496 Diffuse, and 653
  Metal-Rich stars, i.e., a total of $N=5722$ stars.

  \begin{figure*}
    \centering
    \includegraphics[width=\linewidth]{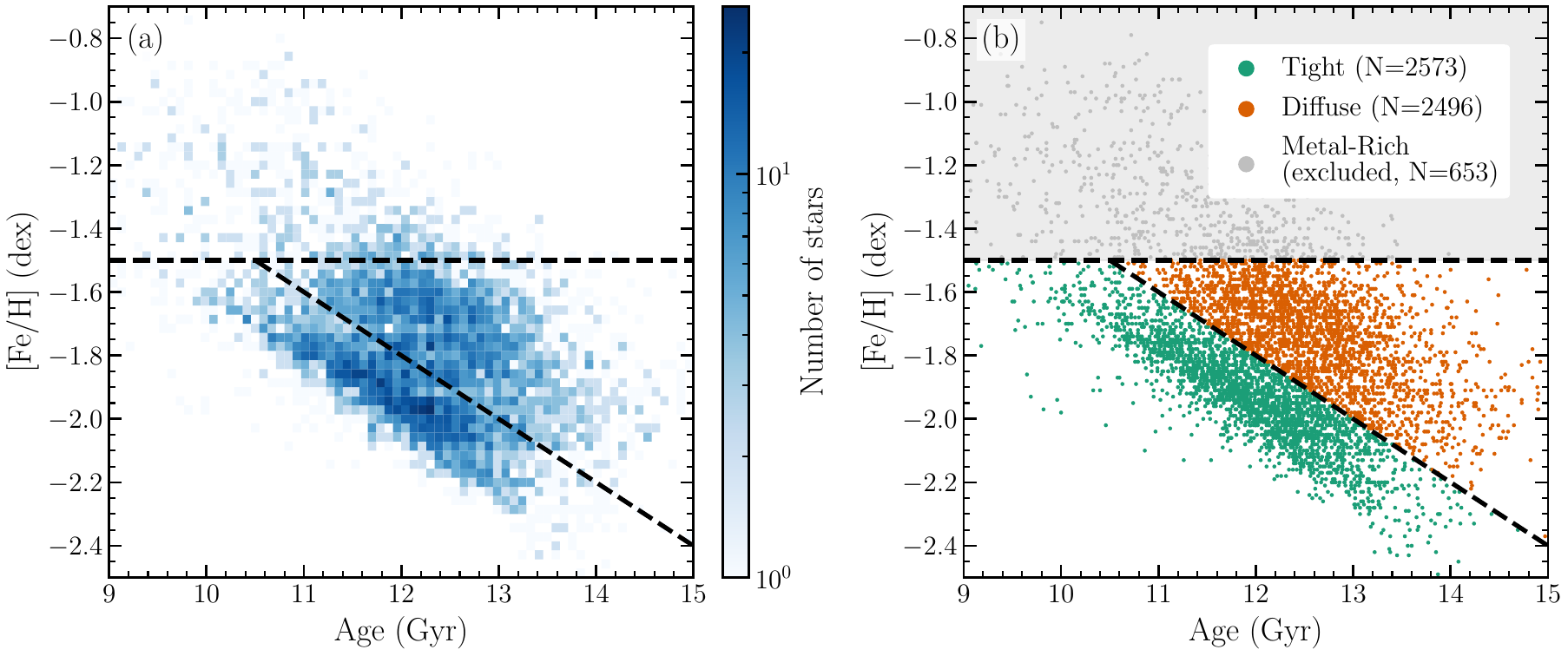}
    \caption{AMR for SGB stars in \ocen, derived by \citet{Clontz2024b} from isochrone fitting to
      HST photometry. (a) AMR colored by the number of stars per bin (log scale), showing the
      distinct sequences present. (b) The three AMR populations (Tight and Diffuse streams from
      \citealt{Clontz2024b}, Metal-Rich group from \citealt{Clontz2026}; see Section~\ref{sec:pop_defs}
      for their use in this work) are indicated: ``Tight'' stars (the younger, narrower sequence at
      fixed metallicity), ``Diffuse'' stars (older at fixed metallicity, broader scatter), and
      ``Metal-Rich'' stars (the most iron-abundant portion, generally younger). Dashed lines mark
      the population boundaries: stars above $\FeH=-1.5$ (gray shading) form the Metal-Rich group,
      which we exclude from the fits, and the diagonal line separates the Tight and Diffuse
      sequences. Median age uncertainties are $\sim$0.68~Gyr.}
    \label{fig:AMR-def}
  \end{figure*}

  \subsubsection{Metallicity Bins}
  The simplest way to define subpopulations is to divide the sample into bins based solely on bulk
  metallicity \FeH.\footnote{The \texttt{spexxy} fits of the MUSE spectra yield total metallicities
    [M/H] \citep{Nitschai2023a}. Throughout this work, we convert these to \FeH\ using the relation of
    \citet{Salaris1993} for $[\alpha/\mathrm{Fe}]=+0.3$, which gives a constant offset,
    $\FeH=[\mathrm{M}/\mathrm{H}]-0.21$, and puts the metallicity bins on the same scale as the AMR
  (Section~\ref{sec:amr_def}).} This approach lets us test whether metallicity alone drives
  kinematic differences among the stars in \ocen. We divide the RGB sample into quintiles of \FeH,
  creating five distinct metallicity bins that span the full metallicity range observed in the
  cluster. In Figure~\ref{fig:MH-pops}, we illustrate the definition of these metallicity bins using
  a kernel density estimate (KDE) of the \FeH\ distribution, indicating the boundaries of each
  quintile with black vertical lines. These stars are required to have $m_{\rm F625W} < 17$ (the RGB
  sample of \citealt{Nitschai2024a} has a mean metallicity uncertainty of 0.04~dex) and a non-null
  \FeH\ measurement, giving $N=8788$; the high-quality-astrometry cut reduces this to $N=7398$, and
  the high-quality LOS cut applied in all analyses (Section~\ref{sec:kinquality}) further reduces it
  to $N=6189$, split into five quintile bins of $\sim$1240 stars each.
  \begin{figure}
    \centering
    \includegraphics[width=\linewidth]{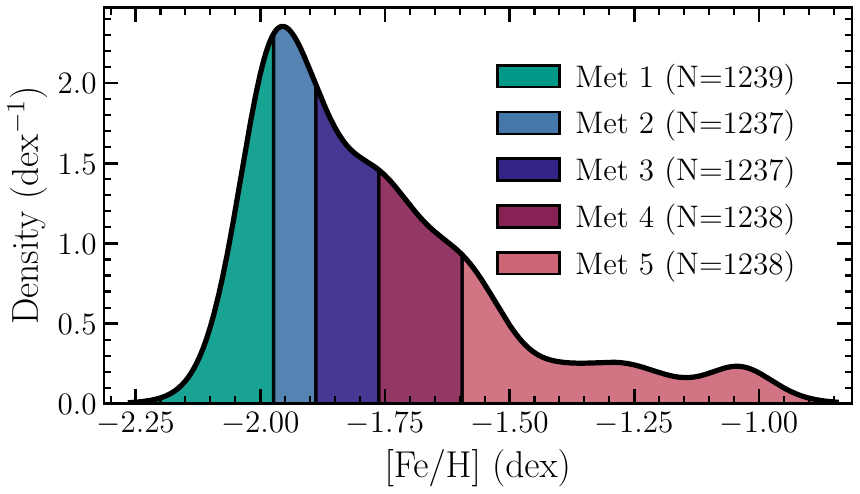}
    \caption{Kernel density estimate (KDE) of the MUSE metallicity \FeH\ distribution for all RGB
      stars used in this work to construct the metallicity bins. The five colored regions indicate
      the quintiles of the distribution, which are used to define the metallicity bins for kinematic
      analysis. The bins are labeled Met~1 (most metal-poor) to Met~5 (most metal-rich). The
      metallicity distribution spans $\sim$1.4~dex and peaks around \FeH$\approx-1.95$, consistent
    with the full metallicity distribution of \ocen as reported by \citet{Nitschai2024a}.}
    \label{fig:MH-pops}
  \end{figure}

  We now model the 2D kinematics of each of these populations (Section~\ref{sec:modeling}).

  \subsection{Kinematic Modeling} \label{sec:modeling}

  We characterize the kinematic structure of each subpopulation using 2D spatial maps of the
  velocity dispersion, rotation, and velocity anisotropy, from which we infer aspects of the
  underlying 3D geometry and identify kinematic differences that may reflect their formation
  histories. The corresponding 1D radial profiles, fit with both discrete (unbinned) and binned
  methods, are presented in Appendix~\ref{app:chromo-1d} as a check for trends and correlations with
  radius.

  \subsubsection{Profiles and Likelihoods}
  \label{sec:profiles}
  We model the velocity dispersion using a Plummer dispersion profile (Equation~\ref{eq:plummer}), which
  describes a centrally peaked dispersion that flattens at large radii and matches our binned
  profiles (Appendix~\ref{app:chromo-1d}). The rotation is modeled as a Lynden-Bell rotation profile
  \citep[e.g.,][]{Lynden-Bell1967, Varri2012, Henault-Brunet2015, Kamann2018}, a rotation curve that
  rises linearly in the core and declines outward, as expected both after violent relaxation
  \citep{Lynden-Bell1967} and for quasi-relaxed rotating clusters \citep{Varri2012}. We implement
  all our models in \texttt{JAX} \citep{jax2018github} and sample the posterior distributions of the
  model parameters using the No-U-Turn Sampler (NUTS) implemented in the \texttt{numpyro} package
  \citep{Phan2019}.

  The Plummer dispersion profile is given by:
  \begin{equation}
    \sigma(R)^2 = \frac{\sigma_0^2}{\sqrt{1 + (R/a)^2}}
    \label{eq:plummer}
  \end{equation}
  where $\sigma_0$ is the central velocity dispersion (km~s$^{-1}$ for LOS fits, mas~yr$^{-1}$ for
  PM fits), $a$ is the scale radius (arcsec), and $R$ is the projected radius (arcsec) from the cluster
  center.

  The Lynden-Bell rotation profile is given by:
  \begin{equation}
    {V_{\rm rot} = \frac{2A_{\rm rot}}{R_{\rm peak}}\cdot\frac{X}{1 + X^2/R_{\rm peak}^2}}
    \label{eq:rotation}
  \end{equation}
  where $A_{\rm rot}$ is the projected peak rotation amplitude (km~s$^{-1}$ or mas~yr$^{-1}$), $R_{\rm peak}$ is the radius (arcsec) at
  which the rotation curve reaches its maximum, and $X$ is the projected distance (arcsec) along the
  kinematic major axis. The model rises linearly at small $X$ and falls as $X^{-1}$ at large $X$.
  When fitting our models to the data, we adopt Gaussian likelihoods, computing the mean velocity
  and velocity dispersion at the location of an individual star. The total likelihood is then the
  product of the likelihoods of all the individual stars, though in practice, we compute the
  log-likelihood and sum over the stars:

  \begin{equation}
    \ln{\mathcal{L}} =
    \sum_{i=1}^N \ln{\left( \frac{1}{\sqrt{2\pi (\sigma_i^2 + \delta_i^2)}}
    \exp{\left( -\frac{(v_i - V_i)^2}{2(\sigma_i^2 + \delta_i^2)} \right)} \right)}
  \end{equation}
  where $v_i$ is the observed velocity of star $i$, $V_i$ is the model mean velocity at the location
  of star $i$, $\sigma_i$ is the model velocity dispersion at that location, and $\delta_i$ is the
  measurement uncertainty for star $i$.

  \subsubsection{Joint 2D Fits and Kinematic Ellipticity}
  \label{sec:joint_fits}
  For each population and velocity component (LOS, $\mu_R$, $\mu_T$), we also construct 2D spatial
  maps of the velocity dispersion, the LOS rotation, and anisotropy ratio
  $\sigma_{\mu_T}/\sigma_{\mu_R}$. This allows us to fit for any elliptical distortion in the
  kinematic fields, which may be blurred or averaged out in the 1D profiles, and constrain the
  azimuthal dependence of the velocity anisotropy. Because the likelihood function $\mathcal{L}$
  evaluates the kinematics at the location of each star, the spatial selection function no longer
  plays a role, allowing us to infer unbiased kinematic fields despite the uneven data coverage from
  the survey footprint. The ellipticity parameter $\epsilon$ is therefore derived purely from the
  kinematic amplitudes that vary as a function of azimuth, not from the spatial distribution of the
  observed stars.\footnote{We do not attempt to measure spatial ellipticity per population: the
    survey selection function is highly non-uniform, leaving under-dense regions that fall directly on
    the cluster minor axis, and exploratory selection-function modeling revealed no significant
  differences between populations.}

  We fit the LOS, $\mu_R$, and $\mu_T$ velocities of each population jointly, with a single
  kinematic axis whose position angle (PA) is free and is set mainly by the LOS
  rotation.\footnote{Fits with an independent dispersion axis for each component give the same
    result: the flattened fields have axes consistent with the rotation axis within the uncertainties,
  and Im and P2 show no coherent flattening signal at any orientation.} Each component has its own
  Plummer profile and ellipticity $\epsilon$ along this axis. We allow $\epsilon$ to take either
  sign (flat prior on $|\epsilon| < 0.95$), with $\epsilon < 0$ describing contours elongated
  perpendicular to the axis. We adopt this somewhat unusual parametrization because a prior bounded
  at $\epsilon=0$ will be biased toward positive $\epsilon$ even when the data are drawn from a
  perfectly spherical distribution. This allows for posteriors that are consistent with zero
  flattening to cluster around $\epsilon=0$, allowing for easier interpretation of the results.

  We measure the evidence for flattening with the Savage--Dickey Bayes factor $B$ \citep{Dickey1971,
  Trotta2008} of the elliptical model against $\epsilon=0$. Because a shallow dispersion profile
  turns a modest azimuthal contrast into a large $\epsilon$, we also quote $\sigma_{\rm
  minor}/\sigma_{\rm major}$ at $R = 200''$, which provides a more intuitive measure of the
  flattening of the velocity field.

  The 2D extension of the Plummer dispersion profile allows for an elliptical distortion by
  replacing the circular radius with an effective radius that accounts for the ellipticity and PA of
  the kinematic field. The 2D dispersion model is then given by:
  \begin{equation}
    \sigma(x,y)^2 = \frac{\sigma_0^2}{\sqrt{1 + R_{\rm eff}(x,y)^2 / a^2}}
  \end{equation}
  where $R_{\rm eff}(x,y) = \sqrt{x^2 + (y / (1 - \epsilon))^2}$ is the effective radius that
  accounts for the ellipticity $\epsilon$ and PA of the kinematic field, where $x$ and $y$ are the
  coordinates aligned with the major and minor axes of the ellipse (rotated from sky coordinates by
    the kinematic position angle PA, as defined in Equations~\ref{eq:x-coord}--\ref{eq:y-coord}
  below), and $\epsilon$ is the kinematic ellipticity.
  \begin{eqnarray}
    x = R \cdot \left( \cos(\theta - \mathrm{PA}) \right)\label{eq:x-coord}\\
    y = R \cdot \left( \sin(\theta - \mathrm{PA}) \right)\label{eq:y-coord}
  \end{eqnarray}

  Likewise, for the rotation field, we allow for an elliptical distortion by replacing the circular
  radius with the effective radius in the rotation model:

  \begin{equation}
    V_{\rm rot}(x,y) = \frac{2 A_{\rm rot}}{R_{\rm peak}} \cdot \frac{x}{1 + R_{\rm eff}(x,y)^2 / R_{\rm peak}^2}
  \end{equation}
  with the same definition of $R_{\rm eff}(x,y)$ as above, but with its own ellipticity,
  $\epsilon_{\rm rot}$; $V_{\rm rot}$ is the projected (LOS) rotation velocity of the cluster. The
  rotation curve of \ocen\ peaks near the edge of our field, so $A_{\rm rot}$ and $R_{\rm peak}$ are
  degenerate in our fits. To compare the rotation of the populations, we therefore evaluate the
  rotation velocity on the kinematic major axis at $R=200''$, $V_{\rm rot}(200'')$, for each
  posterior draw.

  In the joint fits the LOS rotation field fixes the direction of the shared axis, which removes the
  PA$\to$PA$+\pi/2$ ambiguity of a signed ellipticity, so we adopt a flat prior on the PA. We adopt
  large, uninformative priors on all other model parameters. In the joint fits we fit a constant
  mean for each PM component, which absorbs any residual offset in the relative proper motions, and
  fit the LOS systemic velocity as a free constant in the same way. We fit no PM rotation term
  because the proper motions measured by HST are relative to the local cluster motion and therefore
  do not capture the motion of the cluster or streaming motion in the tangential direction, only the
  dispersion about the local mean motion.

  To validate that the kinematic signatures recovered by our model are not artifacts of the
  parametric fitting approach, we compare our anisotropy maps to the non-parametric Voronoi-binned
  maps of \citet{Haberle2025}, using the same sets of high-quality stars. Our models produce maps
  that are in good agreement with the Voronoi-binned maps, indicating that the kinematic features we
  identify are not dependent on the specific modeling approach.\footnote{In particular, we verify
    that the residuals are symmetric about zero with the expected scatter of $\sim 1 \sigma$ given the
    uncertainties in the Voronoi-binned maps, and that there are no spatially coherent residuals that
  would indicate a systematic bias in our modeling approach.}
  \section{Results} \label{sec:results}

  We present the kinematic results for the various subpopulation definitions outlined in
  Section~\ref{sec:pop_defs}. We begin with the chromosome-map populations, which reveal the most
  distinct kinematic differences. We then show that populations defined by pure metallicity bins
  recover no comparable structure, and describe why the AMR sample, which has limited radial
  coverage, cannot test for the P1 flattening.

  \begin{figure*}
    \centering
    \includegraphics[width=\linewidth]{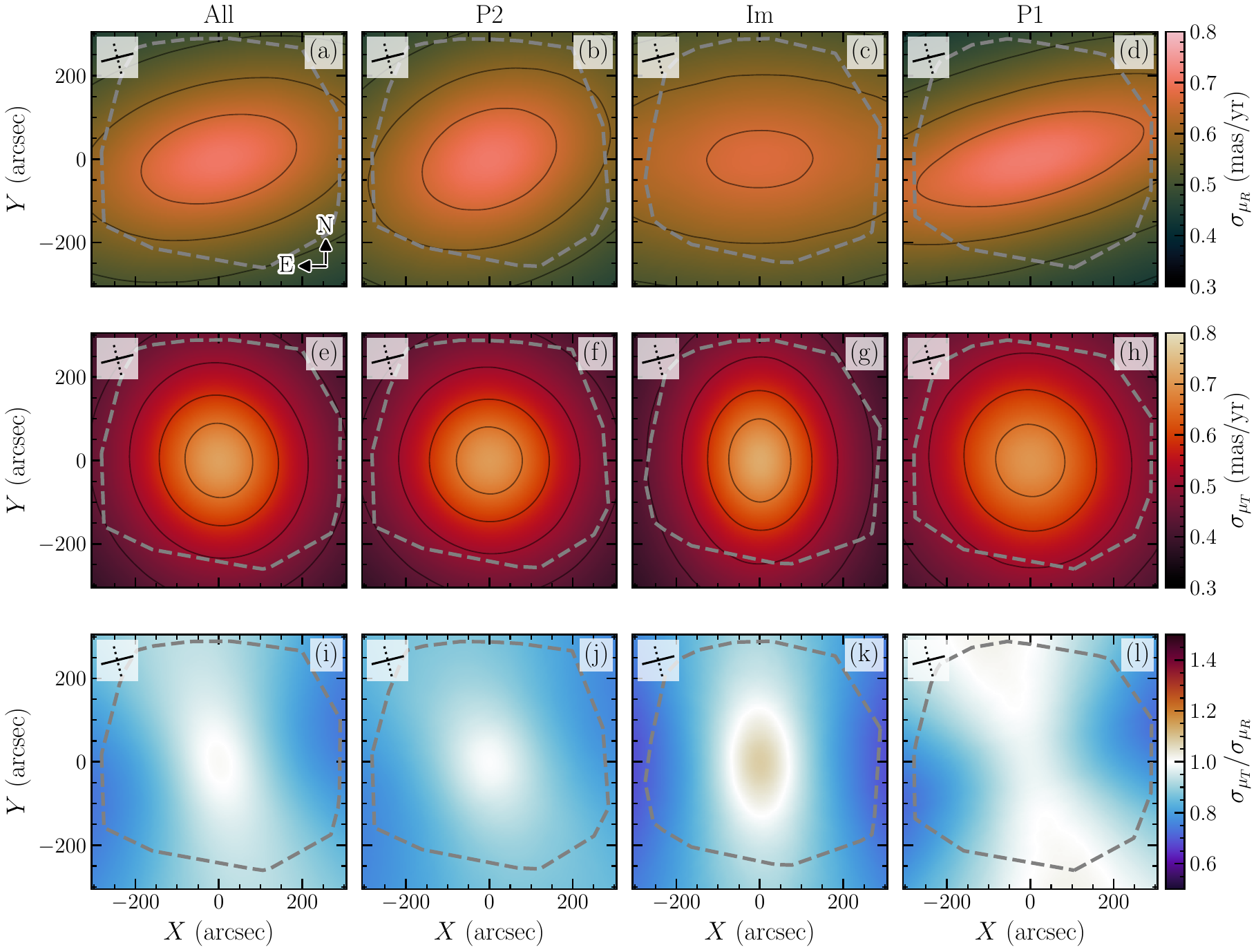}
    \caption{HST proper-motion kinematic maps for the bulk chromosome-map sample (All), P2, Im, and
      P1 (columns, left to right). \textit{Rows, top to bottom:} radial PM dispersion
      $\sigma_{\mu_R}$, tangential PM dispersion $\sigma_{\mu_T}$, and anisotropy ratio
      $\sigma_{\mu_T}/\sigma_{\mu_R}$. The dashed regions indicate the convex hull containing the
      data in each subset. The inset at the top left of each panel marks the cluster major (solid)
      and minor (dotted) axes. Thin black lines in the dispersion panels are iso-dispersion
      contours, evenly spaced across the color range. The anisotropy maps are the ratio of the
      fitted $\sigma_{\mu_T}$ and $\sigma_{\mu_R}$ fields in the rows above. For P1 (panel l), the
      flattened $\sigma_{\mu_R}$ and round $\sigma_{\mu_T}$ fields combine into a quadrupole
      pattern: near-isotropic along the minor axis and radially anisotropic
      ($\sigma_{\mu_T}/\sigma_{\mu_R}\approx0.8$) along the major axis (see
      Section~\ref{sec:res_chromo}). Im and P2 anisotropy maps show no extended quadrupole and are
      near-isotropic (P2) or mildly tangential (Im) in the core, becoming radially anisotropic at
    larger radii in all directions.}
    \label{fig:chromo-maps}
  \end{figure*}

  \begin{figure*}
    \centering
    \includegraphics[width=\linewidth]{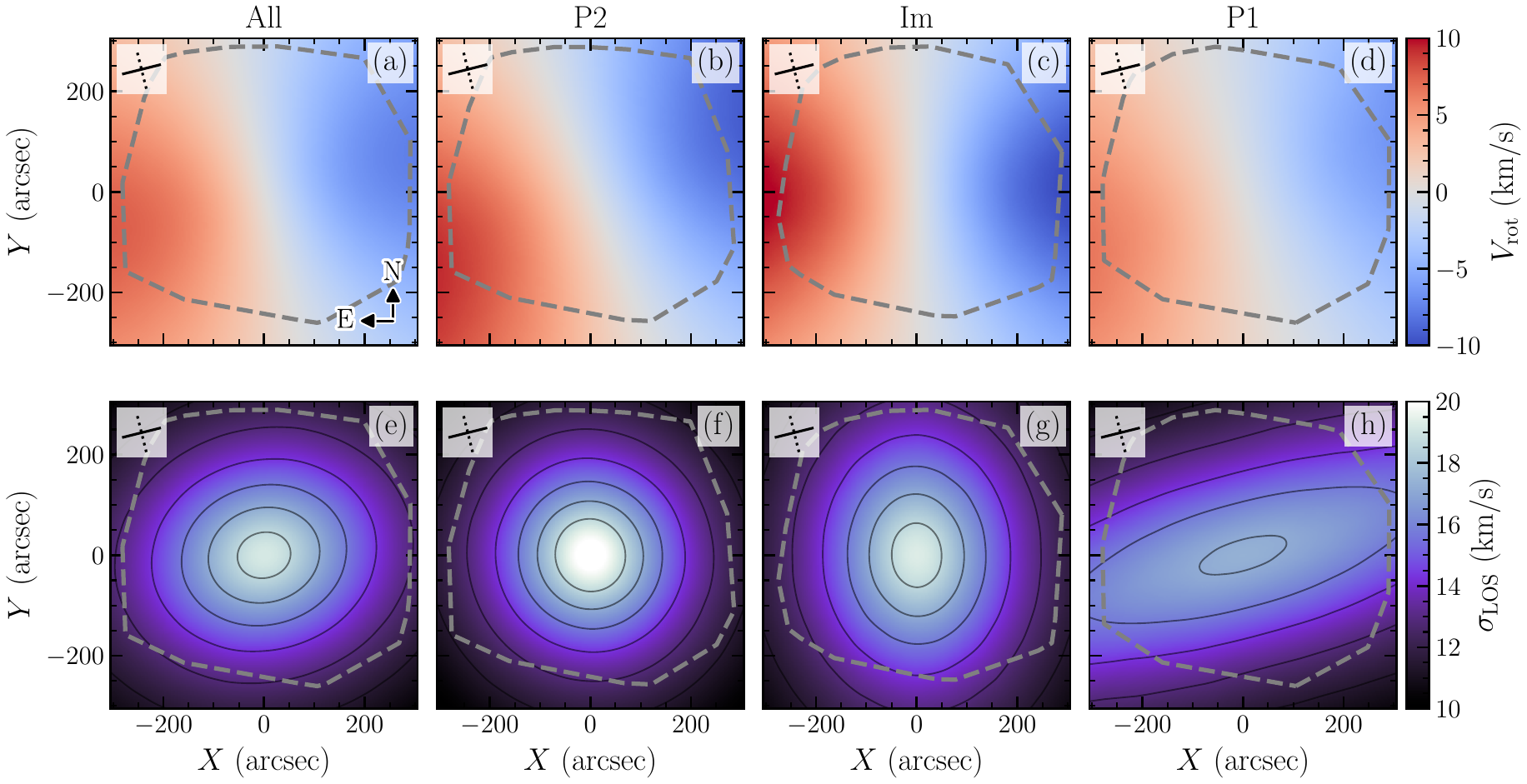}
    \caption{2D MUSE LOS kinematic maps for the bulk chromosome-map sample (All), {P2, Im, and P1}
      (columns, left to right). \textit{Rows, top to bottom:} LOS rotation velocity $V_{\rm rot}$
      and LOS velocity dispersion $\sigma_{\rm LOS}$. Dashed lines indicate the convex hull
      containing the data in each subset. The inset at the top left of each panel marks the cluster
      major and minor axes, as in Figure~\ref{fig:chromo-maps}. Thin black lines in the dispersion
      panels are iso-dispersion contours, evenly spaced across the color range. The P1 dispersion
      field (panel h) is elliptical ($\epsilon \approx 0.7$; see Section~\ref{sec:res_chromo}). The Im
      and P2 dispersion fields are consistent with azimuthally symmetric distributions.}
    \label{fig:chromo-los-maps}
  \end{figure*}

  \begin{figure*}
    \centering
    \includegraphics[width=\linewidth]{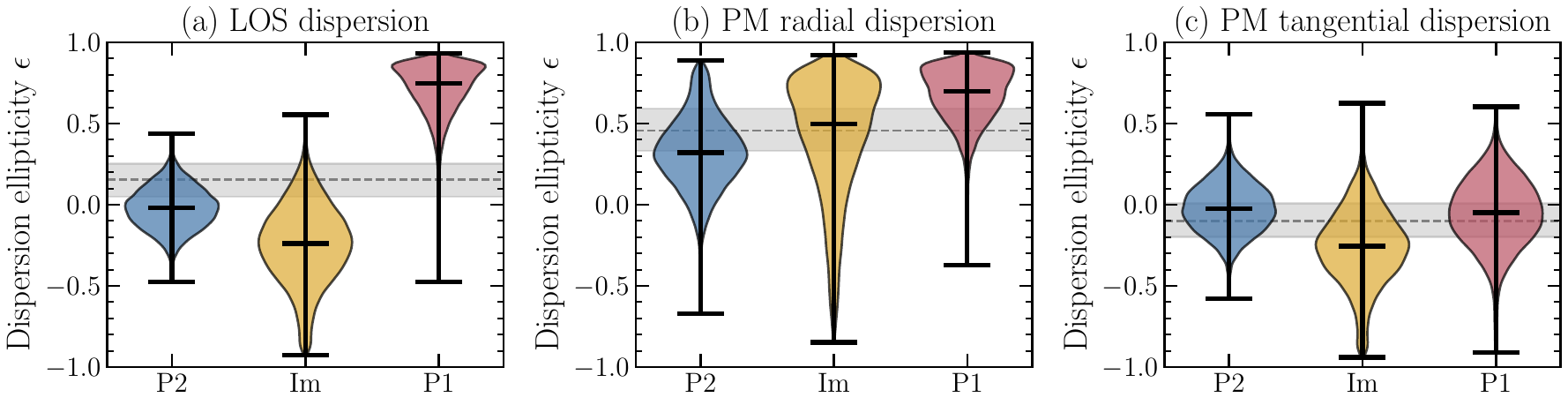}
    \caption{Posterior distributions of the dispersion-field ellipticity $\epsilon$ for the P2, Im,
      and P1 chromosome-map populations, from the elliptical fits to (a) $\sigma_{\rm LOS}$, (b)
      $\sigma_{\mu_R}$, and (c) $\sigma_{\mu_T}$, compared against the bulk chromosome-map sample
      posterior (median as dashed line, 68\% interval as gray band). Horizontal bars mark each
      posterior's median and full range. Negative $\epsilon$ describes a dispersion field elongated
      perpendicular to the kinematic axis (Section~\ref{sec:modeling}). P1 is flattened in
      $\sigma_{\rm LOS}$ and $\sigma_{\mu_R}$ (Bayes factors $B\approx50$ and 30), while Im and P2
      are consistent with $\epsilon=0$ in the LOS. In $\sigma_{\mu_R}$, the bulk sample is also
      flattened ($B\approx40$) and the Im and P2 posteriors include $\epsilon=0$. In
      $\sigma_{\mu_T}$, all posteriors include $\epsilon=0$. The LOS rotation posteriors are shown
      in Figure~\ref{fig:rotation-violin}.}
    \label{fig:chromo-ellipticity-violin}
  \end{figure*}

  \subsection{Chromosome Map Populations}
  \label{sec:res_chromo}

  The chromosome-map populations show the clearest differences in kinematic structure
  (Figures~\ref{fig:chromo-maps} and \ref{fig:chromo-los-maps}).
  Figure~\ref{fig:chromo-ellipticity-violin} shows the posterior distributions of the elliptical
  fits directly: in the LOS, the P1 ellipticity posterior is offset from zero, from the bulk, and
  from the P2/Im posteriors. In $\sigma_{\mu_R}$, P1 is again the most flattened population and is
  offset above the bulk, but its posterior overlaps those of Im and P2, which are consistent with
  the bulk sample. The PM contrast is smaller because the whole cluster is flattened in
  $\sigma_{\mu_R}$ ($B\approx40$) but not in $\sigma_{\rm LOS}$: the cluster rotation, seen in
  projection, flattens the $\sigma_{\mu_R}$ field of every population~\citep{Haberle2025}, so the
  additional flattening of P1 is measured against an already flattened baseline, whereas in the LOS
  the bulk, Im, and P2 are close to round. In 1D, the Im and P2 populations display largely similar
  dispersion and rotation profiles (Appendix~\ref{app:chromo-1d}), while the P1 stars exhibit a
  shallower LOS dispersion profile (Figure~\ref{fig:chromo-los}); the azimuthally averaged PM
  anisotropy profiles of all three populations agree within the uncertainties
  (Figure~\ref{fig:chromo-pm}).

  Within $r_h$, P1 rotates no faster than Im and P2, and somewhat more slowly in projection
  (Figures~\ref{fig:chromo-los-maps} and \ref{fig:chromo-los}).  At $200''$, the elliptical fits
  give $V_{\rm rot}(200'')=5.0\pm0.8$~km~s$^{-1}$ for P1, against $7.2\pm0.6$ (P2) and
  $8.5\pm1.0$~km~s$^{-1}$ (Im) (Figure~\ref{fig:rotation-violin}d). The peak amplitudes $A_{\rm
  rot}$ overlap (Figure~\ref{fig:rotation-violin}a), because the peak of the rotation curve lies at
  the edge of our field and $A_{\rm rot}$ trades off against $R_{\rm peak}$
  (Section~\ref{sec:modeling}).  The axis of rotation, however, may differ for Im. The elliptical
  fits place its kinematic position angle (the line of maximum rotation) at PA $=92\pm6^\circ$,
  against $106\pm5^\circ$ for P1 and $112\pm5^\circ$ for P2 (Figure~\ref{fig:rotation-violin}j).
  Fitting the individual LOS velocities of each population with only a constant and a dipole in
  position angle, $v = c + a\cos\theta + b\sin\theta$, which assumes no rotation curve, dispersion
  profile, or flattening, places the Im axis $20\pm8^\circ$ from that of P2. Because the rotation
  signal grows with radius, both measurements are set mostly by the outer stars.\footnote{Splitting
    each population at the median radius of the sample ($\approx160''$), the inner Im and P2 axes
    agree ($98\pm10^\circ$ and $97\pm9^\circ$), whereas the outer axes differ by $23\pm8^\circ$
  ($90\pm6^\circ$ and $113\pm5^\circ$).}
  
  In 2D, the kinematic maps reveal clear differences in the shape of the dispersion fields. When we
  consider the azimuthal dependence and flattening of the dispersion fields, the P1 stars exhibit
  kinematic flattening along the major axis in $\sigma_{\rm LOS}$ and $\sigma_{\mu_R}$
  ($\epsilon_{\rm LOS} = 0.75^{+0.11}_{-0.19}$ and $\epsilon_{\mu_R} = 0.70^{+0.14}_{-0.19}$, with
    Bayes factors $B \approx 50$ and $30$; Figure~\ref{fig:chromo-ellipticity-violin}, with maps in
  Figures~\ref{fig:chromo-maps} and \ref{fig:chromo-los-maps}), compared to the spherical dispersion
  fields of the Im and P2 stars (Figures~\ref{fig:chromo-los-maps} and
  \ref{fig:chromo-ellipticity-violin}).

  For P1, the elongated $\sigma_{\mu_R}$ field and the round $\sigma_{\mu_T}$ field combine into a
  quadrupole pattern: $\sigma_{\mu_T}/\sigma_{\mu_R}\approx1$ along the minor axis and $\approx0.8$
  along the major axis. The Im and P2 stars show no extended quadrupole: their maps are
  near-isotropic (P2) or mildly tangential (Im, $\sigma_{\mu_T}/\sigma_{\mu_R}\approx1.1$) in the
  core and become radially anisotropic beyond $\sim100''$ in all directions. \citet{Haberle2025}
  identified quadrupole-like ``plumes'' of the same form in the PM anisotropy map of the whole
  cluster (their Figure~11), which they attribute to the physical anisotropy combined with the
  projection of the cluster rotation.

  The P1 ellipticities correspond to $\sigma_{\rm minor}/\sigma_{\rm major} = 0.81\pm0.05$ in both
  components at $R=200''$, consistent with the $\sim$15\% contrast we measure without a parametric
  model.\footnote{We also compared the dispersion in $\pm30^\circ$ wedges about the major and minor
    axes, in three radial bins. P1 gives $\sigma_{\rm minor}/\sigma_{\rm major}=0.87\pm0.05$ (LOS) and
    $0.83\pm0.04$ ($\mu_R$), against 0.94--1.02 for Im and P2 and $0.98$--$0.99\pm0.04$ for round mock
    datasets with the P1 positions and uncertainties, confirming the signal seen in the parametric
  models.} At the same radius, Im and P2 give $1.10^{+0.12}_{-0.09}$ and $1.01^{+0.06}_{-0.05}$ in
  the LOS and $0.92\pm0.08$ and $0.92\pm0.05$ in $\mu_R$, and the bulk sample $0.94\pm0.04$ and
  $0.88^{+0.04}_{-0.03}$. Im and P2 show no significant flattening in any component ($B \leq 1.4$;
  Figure~\ref{fig:chromo-ellipticity-violin}). Kinematic ellipticity is independently detected in
  both $\sigma_{\rm LOS}$ and $\sigma_{\mu_R}$, as these two components are measured from
  independent datasets, HST astrometry and MUSE spectroscopy, though in the same sample of stars
  (Figure~\ref{fig:chromo-ellipticity-violin}).

  We interpret the apparent difference between the slightly shallower 1D dispersion profile
  (Figure~\ref{fig:chromo-los}) and the elliptical 2D field (Figure~\ref{fig:chromo-los-maps}) as a
  result of the data footprint, which is elongated and nearly aligned with the cluster geometry: the
  majority of stars at larger radii fall along the flattened direction, which would bias the 1D
  profiles toward higher dispersion at large radii and thus a shallower radial gradient than the 2D
  field.

  In summary, the chromosome map isolates a single dynamically distinct component, the primordial
  stars, which show a flattened dispersion field in both datasets
  (Figure~\ref{fig:chromo-ellipticity-violin}), with the PM anisotropy quadrupole that the
  $\sigma_{\mu_R}$ flattening implies (Figure~\ref{fig:chromo-maps}), and rotation no faster than
  that of Im and P2 within $r_h$ (Figure~\ref{fig:chromo-los-maps}). Having established that the
  chromosome map does reveal dynamically distinct kinematic components, we next test whether simpler
  metallicity-only divisions recover any comparable structure.

  \subsection{Metallicity Bins}
  \label{sec:res_metallicity}

  We show the definition of the five \FeH\ quintiles in Figure~\ref{fig:MH-pops} and their
  ellipticity posteriors in Figure~\ref{fig:MH-ellipticity-violin}. The 1D LOS and PM profiles are
  presented in Figures~\ref{fig:MH-los} and \ref{fig:MH-pm} (Appendix~\ref{app:metallicity-bins}).
  The metallicity quintiles show consistent LOS and PM dispersion profiles across all bins
  (Figures~\ref{fig:MH-los} and \ref{fig:MH-pm}), no discernible trend in rotation with metallicity
  (Figure~\ref{fig:MH-los}), and PM anisotropy curves that agree within the uncertainties
  (Figure~\ref{fig:MH-pm}). None of the quintile ellipticity posteriors is a secure detection, and
  they show no coherent trend with \FeH. The only secure flattening we recover is when all
  populations are combined. Figure~\ref{fig:MH-ellipticity-violin} shows this directly: the
  ellipticity posteriors of the individual \FeH\ quintiles overlap both each other and the
  bulk-cluster posterior in both the LOS and PM fits.

  We interpret this as following directly from the substantial overlap in the \FeH\ distributions of
  P1, Im, and P2 \citep{Clontz2026, Wang2026a}: each \FeH\ bin contains a mixture of the
  kinematically distinct chromosome-map populations, and therefore dilutes their signatures.

  In summary, we find no coherent departure from the bulk cluster in ellipticity, rotation, or
  anisotropy for any of the five \FeH\ quintiles, and no trend with \FeH\
  (Figure~\ref{fig:MH-ellipticity-violin}). We next test the more structured AMR divisions.

  \begin{figure*}
    \centering
    \includegraphics[width=\linewidth]{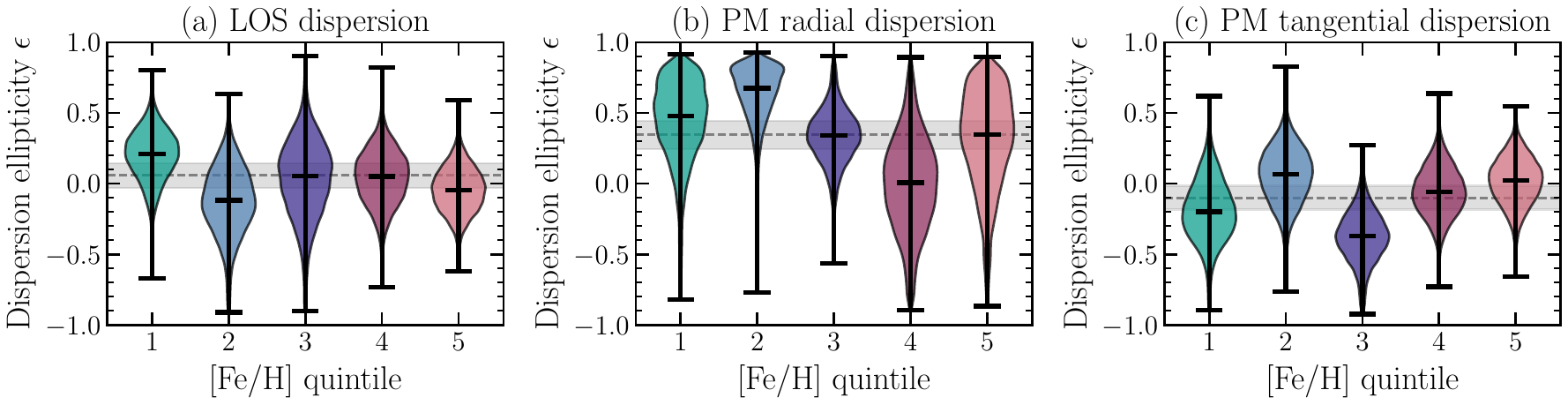}
    \caption{As in Figure~\ref{fig:chromo-ellipticity-violin}, but for the \FeH\
      quintiles, compared against the bulk \FeH\ sample posterior. Unlike the chromosome-map
      decomposition, no \FeH\ quintile is preferentially offset from the bulk posterior in any
    component.}
    \label{fig:MH-ellipticity-violin}
  \end{figure*}

  \subsection{Age--Metallicity Populations}
  \label{sec:res_amr}

  The AMR sample ($N_{\rm Tight}=2573$ and $N_{\rm Diffuse}=2496$; we exclude the small, $N=653$,
  Metal-Rich sequence) is limited to the central region with reliable F606W coverage: ninety percent
  of its stars lie within $139''$ of the center, inside the core radius ($R_{\rm core}=4.3$~pc
  $\approx163''$; \citealt{Baumgardt2018}, updated online catalog). Inside the core the dispersion
  field is nearly flat, so it constrains the ellipticity only weakly, and every AMR posterior,
  including that of the bulk AMR sample, is broad (Figure~\ref{fig:amr-ellipticity-violin}). With
  this coverage, the AMR populations do not allow a meaningful comparison of kinematic structure
  with the chromosome-map or metallicity divisions. We show their ellipticity posteriors and 1D
  profiles in Appendix~\ref{app:amr}.


  \section{Discussion} \label{sec:discussion}

  \begin{figure*}[t]
    \centering
    \includegraphics[width=0.85\linewidth]{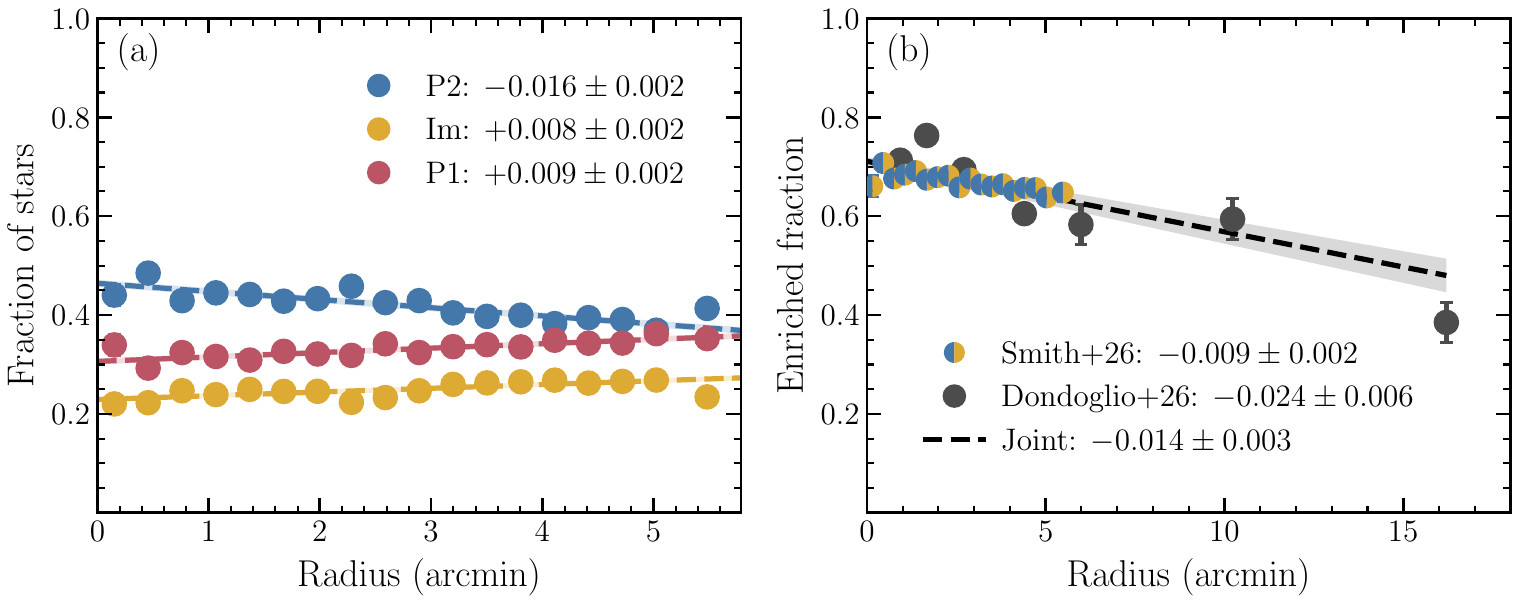}
    \caption{Fraction of P1 (red), Im (yellow), and P2 (blue) stars as a function of projected
      radius, using \citet{Clontz2026}'s stacked, data-driven clustering sample. P2 is the dominant
      population in the cluster core and its fraction declines outward, while the P1 fraction rises
      with radius, as does that of Im. (a) Fractions of each population in radial bins; dashed lines
      are linear fits and shaded bands their $1\sigma$ intervals. (b) The enriched fraction (P2+Im)
      from this work (two-color symbols) compared with the enriched (2P) fraction of
      \citet{Dondoglio2026} (gray; their Figure 5), which they compute relative to their canonical
      1P+2P stars only, excluding their iron-rich AI and AII groups, and whose outermost bin is
      centered at $\sim16\arcmin$ and reaches $\sim20\arcmin$; the black dashed line and gray band
      show a joint linear fit to both datasets. Slopes in the legends are in fraction per arcmin;
    the P2, P1, and Im slopes differ from zero at $7.0\sigma$, $4.3\sigma$, and $4.2\sigma$.}
    \label{fig:radial-fractions-chromo}
  \end{figure*}

  \subsection{The P1/P2 dichotomy}
  \label{sec:dichotomy_compiled}

  Within \ocen's half-light radius, the chromosome-map primordial (P1) and enriched (Im, P2)
  populations are dynamically distinct (Section~\ref{sec:res_chromo}). P1's dispersion field is
  flattened along the cluster major axis, with kinematic ellipticity $\epsilon\approx0.7$ recovered
  independently from HST proper motions and MUSE LOS velocities
  (Figure~\ref{fig:chromo-ellipticity-violin}), and because its $\sigma_{\mu_T}$ field is round, its
  proper-motion anisotropy follows a quadrupole pattern (Figure~\ref{fig:chromo-maps}),
  near-isotropic along the minor axis and radially anisotropic along the major axis. Im and P2 are
  consistent with spherical symmetry in the LOS dispersion field and are no more flattened than the
  bulk sample in the proper motions (Section~\ref{sec:res_chromo}), show no extended quadrupole, and
  transition from near-isotropy or mild tangential anisotropy in the core to radial anisotropy at
  intermediate radii. Because P1 rotates no faster than P2/Im within $r_h$, and more slowly at
  $200''$, while its LOS flattening and extended quadrupole appear in P1 alone, these features are
  not a projection of the global cluster rotation. Neither the flattening nor the quadrupole appears
  when the cluster is divided by metallicity (Section~\ref{sec:res_metallicity}), so the
  star-by-star abundance variations traced by the chromosome map, not bulk metallicity, separate the
  dynamically distinct components of \ocen. \FeH\ and AMR position may trace gas supplied by the
  host galaxy \citep{Laporte2026}, whereas the chromosome map traces enrichment within the cluster.
  Divisions by \FeH\ or by position on the AMR mix the chromosome-map populations: among stars with
  both AMR and chromosome-map labels from \citet{Clontz2026}, the Tight AMR sequence is $\sim$21\%
  P1, 34\% Im, and 45\% P2, and the Diffuse sequence $\sim$47\%, 20\%, and 33\%. The Diffuse
  sequence is more P1-rich than the bulk chromosome-map sample, but the AMR sample covers only the
  core and cannot test whether this mixture carries the P1 flattening (Section~\ref{sec:res_amr}).

  We find that P2 is also more centrally concentrated than P1: in \citet{Clontz2026}'s stacked,
  data-driven clustering sample (Figure~\ref{fig:radial-fractions-chromo}), the P2 fraction declines
  with projected radius while the P1 fraction rises, as seen even more clearly in studies covering a
  larger radial range \citep[e.g.,][]{Bellini2009, Scalco2024c, Dondoglio2026}. Because relaxation
  would only act to erase this difference, the observed radial trend indicates that P2 formed more
  centrally concentrated than P1, and that the two populations have not yet fully mixed. P2 was
  probably also denser than P1 when it formed, as most formation scenarios assume
  \citep[e.g.,][]{Aros2025a}, and is now relaxing toward a more extended configuration.

  The main difference we detect lies in the distribution of orbits. In the self-consistent rotating
  models of \citet{Varri2012}, the density and dispersion fields are flattened mainly by rotation,
  and a faster-rotating model is a flatter one. P1 is the flatter component despite rotating more
  slowly, which suggests that the two components differ in orbital structure. Within $r_h$ we
  measure the shape of this structure, a flattened dispersion field. The flattening is consistent
  with P1 orbits whose angular-momentum vectors are more strongly aligned with the rotation axis.
  Recovering the 3D orbit distribution requires the dynamical modeling we outline in
  Section~\ref{sec:conclusions}.

  Detecting the P1 flattening requires azimuthal in addition to radial coverage: it is an
  intrinsically 2D signal that a purely radial $\sigma(r)$ profile would miss or dilute if the data
  under-sample one range of position angles. The combination of deep HST proper motions and
  wide-field MUSE LOS velocities in oMEGACat provides the angular coverage this measurement needs.
  
  The decomposition into P1, Im, and P2 may, in addition to the projected rotation
  (Section~\ref{sec:res_chromo}), partly explain the mild flattening of the cluster's proper-motion
  dispersion field reported by \citet{Haberle2025}, $1 - \sigma_y/\sigma_x = 0.09$ from a 2D
  Gaussian fit to the Voronoi-binned dispersion map. Their value is a dispersion ratio, not an
  ellipticity: on a nearly flat dispersion profile, a small azimuthal contrast gives a large
  $\epsilon$ (Section~\ref{sec:joint_fits}). The comparable quantity for P1 is $1-\sigma_{\rm
  minor}/\sigma_{\rm major}\approx0.19$ (Section~\ref{sec:res_chromo}). P1 makes up about 30\% of
  our chromosome-map sample (Appendix~\ref{app:cuts}), so if Im and P2 are round, the full sample
  would show a contrast of $\approx0.06$, comparable to theirs. Our bulk samples, which mix all
  three populations, give small values in the LOS ($\epsilon=0.06$--$0.16$; the smaller, central AMR
  sample is unconstrained) and, for the two larger samples, in the tangential proper motions
  ($\epsilon_{\mu_T}=-0.10$, consistent with zero), and only their radial proper-motion dispersion
  is more flattened ($\epsilon_{\mu_R}=0.35$--$0.46$).
  
  Because Im and P2 share similar kinematics, we treat them as a single enriched component (P2/Im)
  throughout, returning to whether they represent one enrichment event or two in
  Section~\ref{sec:im_p2}. Their radial distributions do differ: in the \citet{Clontz2026} sample,
  P2 is more centrally concentrated than Im (median radii of $175''$ and $187''$), whereas Im is
  distributed like P1 (median $185''$; Figure~\ref{fig:radial-fractions-chromo}).

  These results constrain any viable formation scenario in three ways. First, P1 must have formed
  in, or later been brought into, a substantially flattened configuration: its kinematic ellipticity
  of $\epsilon\approx0.7$ requires either a birth environment with substantial angular momentum or
  geometric flattening, or a later event that induced the flattening dynamically, such as a merger
  or assembly event in which P1 fell in on a tangential orbit
  (Section~\ref{sec:hierarchical_assembly}). Second, P1 must rotate no faster than P2/Im within
  $r_h$, and more slowly at $200''$. Third, P2/Im have a dispersion field consistent with spherical
  symmetry today, despite \ocen's long relaxation time, so they either formed round or became round
  early, while they were more compact than today, as their central concentration also suggests.

  \subsection{Helium abundances in the chromosome-map populations}
  \label{sec:helium}

  Many previous studies of the spatial and kinematic properties of \ocen have focused on the
  cluster's multiple main sequences, splitting stars into He-normal and He-rich groups
  \citep{Sollima2007,Bellini2009,Bellini2018}. The chromosome-map populations give us the closest
  equivalent of these populations on the RGB: at fixed \FeH, P2 is He-enhanced by $\Delta
  Y\approx0.11$--$0.15$ relative to P1, and Im only mildly ($\Delta Y\approx0.03$;
  \citealt{Clontz2025a}). On the main sequence, the He-rich stars can be further split into the bMS,
  MSd, and MSa \citep{Bellini2017a}. All three groups are He-enriched, but they differ in
  metallicity, from the bMS to the most metal-rich MSa. \citet{Clontz2026} map their subpopulations
  to the labels of \citet{Bellini2017a}: by number, the P2 stars are a mixture of the bMS
  ($\sim$74\%), MSd ($\sim$18\%), and MSa ($\sim$8\%), so P2 is therefore not a homogeneous
  population, though it contains only the He-rich groupings.

  The spread of He within P2 is unlikely to affect our main results. The Im and P2 populations
  differ in He enhancement by a factor of $\sim$4--5 yet have similar kinematics within $r_h$, and
  at larger radii the MSd- and bMS-like groups of \citet{Bellini2018} are both radially anisotropic,
  suggesting that the He-rich groups move similarly. We cannot exclude, however, that an outlier
  He-rich group with distinct kinematics is mixed into P2 and diluted by the other He-rich stars. On
  the MS, where these groups can be separated directly, such a group could be
  isolated (Section~\ref{sec:conclusions}). The main signal we detect, the kinematic flattening of
  P1, is in the He-normal stars and is unaffected by the composition of P2.

  \subsection{Agreement with dynamical models and larger-radius studies}
  \label{sec:dynamical_context}
  \label{sec:larger_radii}

  The signatures we observe in the inner regions are consistent with results at larger radii. At
  $\sim3.5\,r_h$, the He-rich MS-II stars of \citet{Bellini2018}, which correspond to our P2, are
  more radially anisotropic than the He-normal MS-I stars, with MS-I rotating faster;
  \citet{Ziliotto2026} (JWST NIRCam plus archival HST) find at $0.9$--$2.3\,r_h$ that their
  lower-stream stars (LS, analogous to our P1) are less radially anisotropic and show proper-motion
  skewness that suggests stronger rotation than the upper-stream stars (US, analogous to our P2/Im),
  with a higher tangential dispersion than the US stars at the same radial dispersion and a larger
  angular momentum dispersion. \citet{Pagnini2026a} extend this comparison to $1$--$40\arcmin$ with
  APOGEE \citep{Majewski2017, Abdurrouf2022} data, where the Al-poor (P1-like) group is
  near-isotropic and the Al-rich (P2/Im-like) group is significantly radially anisotropic. Together,
  these studies show the primordial stars differing from the enriched stars outside $r_h$ in three
  ways: a larger angular-momentum dispersion at $0.9$--$2.3\,r_h$ \citep{Ziliotto2026}, weaker
  radial anisotropy out to large radii, where the two groups nonetheless rotate alike
  \citep{Pagnini2026a}, and faster rotation at $\sim3.5\,r_h$ \citep{Bellini2018}. Within $r_h$,
  where the cluster rotation curve flattens beyond $\sim150''$ \citep{Haberle2025} and our fitted
  peak lies near the edge of the field (Section~\ref{sec:res_chromo}), P1 instead rotates more
  slowly than P2/Im (Section~\ref{sec:res_chromo}). P1 differs there in its orbital structure: a
  flattened dispersion field, consistent with angular-momentum vectors more strongly aligned with
  the rotation axis (Section~\ref{sec:dichotomy_compiled}). The differentially rotating equilibrium
  models of \citet{Varri2012} are flattened most strongly in their central regions and are nearly
  round in their outer parts, so the P1 flattening need not persist to larger radii. The faster
  rotation of P1 at $\sim3.5\,r_h$ can arise in two ways. P1 may have formed with more angular
  momentum than P2/Im. Alternatively, the initially compact enriched stars may have diffused outward
  (\citealt{Tiongco2019}; Section~\ref{sec:relaxation}). A P1 born in a rotating, flattened
  environment would account for both its inner flattening and its faster outer rotation, but the
  present data do not exclude the outward diffusion of the enriched stars.

  \subsection{Relaxation and Dynamical Evolution}
  \label{sec:relaxation}
  Cospatial dynamical evolution can help explain the radial trend we observe. Two-body relaxation
  erases the differences between populations at different rates for different quantities, which is
  why some of the differences we observe survive while others may not. \citet{Tiongco2019} simulate
  an initially compact P2 population embedded in an extended P1, with the P2 rotating faster in the
  inner regions: two-body relaxation matches the rotation curves first in the inner regions, well
  before anisotropy differences are erased or spatial mixing completes. The P2 stars that diffuse
  outward rotate more slowly than P1 in the outermost regions, so P2 rotates faster over most of the
  cluster and P1 only in the outskirts, the pattern that our data and those of \citet{Bellini2018}
  show across the two radial ranges (Section~\ref{sec:larger_radii}). \citet{Aros2025a} confirm in
  their $N$-body models that anisotropy differences survive to $n_h\gtrsim6$ (where $n_h$ is the
  number of elapsed half-mass relaxation times), long after concentration differences decay. Neither
  set of models is isolated, but both use a static tidal field: a circular orbit in a point-mass
  host \citep{Tiongco2019} or a tidal-radius truncation \citep{Aros2025a}. Neither includes the
  time-varying tides or disk shocks that \ocen\ experiences on its eccentric orbit, which could heat
  the more extended P1 more strongly than the compact P2/Im.

  \ocen's present-day half-mass relaxation time ($\sim$21~Gyr; \citealt{Baumgardt2018,
  Baumgardt2021}, updated online catalog) is long enough that one might expect even more pronounced
  differences than we observe. If, however, \ocen expanded once its tidal field weakened on
  accretion into the MW halo (as \citealt{Giersz2025a} model for NGC~3201), the relaxation time
  would have been shorter in the past, and the populations may have mixed more than the present-day
  value implies.

  Independently, P2's helium enrichment \citep{Clontz2025a} makes its giants less massive than those
  of P1 ($0.82\,M_\sun$ for P1 vs.\ $0.62\,M_\sun$ for P2; \citealt{Wang2025b}), because helium-rich
  stars evolve faster. The younger age of P2 works in the opposite direction, but this effect is
  much smaller. From equipartition alone ($\eta\lesssim0.09$; \citealt{Haberle2025}), this mass
  difference predicts a dispersion difference of at most $\sim$0.5~km~s$^{-1}$, several times
  smaller than the $\sim$3--4~km~s$^{-1}$ difference between the major- and minor-axis dispersions
  of P1. A mass difference also changes the dispersion equally in all directions, so it cannot
  flatten the dispersion field, and equipartition cannot produce the kinematic differences we
  measure.

\subsection{Comparison with Typical Globular Clusters}\label{sec:typical-gc-comparison} The
kinematic differences we observe in \ocen\ are similar to many of those found in typical globular
clusters. There, the enriched stars are often, though not always, more centrally concentrated
\citep[e.g.,][]{Dalessandro2019, Leitinger2023a}, and they rotate faster than the primordial stars
\citep{Dalessandro2024a} or at a similar rate \citep{Milone2018b, Martens2023a, Leitinger2025}. They
become more radially anisotropic outside the core, while the primordial stars remain closer to
isotropy \citep{Libralato2023a, Dalessandro2024a, Cordoni2025a}. These differences are strongest in
dynamically young clusters and fade with dynamical age \citep{Dalessandro2024a, Cordoni2025a}, and
they match models in which the enriched stars form in the cluster core: such stars start centrally
concentrated with faster rotation \citep{Henault-Brunet2015, Lacchin2022}. NGC~2808, which also
hosts strongly He-enhanced populations \citep{Piotto2007}, shows a similar pattern: its He-rich
stars become radially anisotropic beyond $\sim1.5\,r_h$ through a lower tangential dispersion, while
inside $r_h$ the populations are indistinguishable \citep{Bellini2015, Griggio2025a}.

  Our enriched stars largely follow this pattern: they are centrally concentrated, radially
  anisotropic outward \citep[see also][]{Bellini2018}, and rotate at least as fast as P1 within
  $r_h$, although at $\sim3.5\,r_h$ the primordial stars rotate faster \citep{Bellini2018}. P1 does
  not: no comparable flattening of the dispersion field has been reported for the primordial stars
  of typical clusters, whereas P1 has the most flattened dispersion field of any population in
  \ocen. The flattened P1 component therefore has no counterpart in typical GCs, which suggests that
  it reflects \ocen's origin as a nucleus rather than the process that produced the enriched stars.

  \subsection{Formation Scenarios}
  \label{sec:hierarchical_assembly}
  \label{sec:self_enrichment}

  \subsubsection{Observational Constraints}
  Any formation scenario for \ocen\ must reproduce six properties of its populations, constrained here
  and in earlier work:
  {
    \begin{itemize}
      \item \textit{Kinematics.} Within $r_h$, P1 is the flattened component and rotates
        2--3$\sigma$ more slowly than P2/Im at $200''$, while P2 and Im are round and become
        radially anisotropic outward (Section~\ref{sec:res_chromo}).
      \item \textit{Concentration.} P2 is more centrally concentrated than P1 and Im, while Im is
        distributed like P1 (Section~\ref{sec:dichotomy_compiled};
        Figure~\ref{fig:radial-fractions-chromo}).

      \item \textit{Ages.} The population ages decrease from P1 ($\approx$12.6~Gyr) through Im
        ($\approx$12.2~Gyr) to P2 ($\approx$11.6~Gyr). P1 is older than P2 at every \FeH, and Im
        falls between P1 and P2 in mean age at every metallicity where it is present
        \citep{Clontz2026}. The most metal-rich stars of \ocen\ formed at $\approx$11~Gyr
        \citep{Clontz2024b}.
      \item \textit{Chemistry.} $[(\mathrm{C+N+O})/\mathrm{Fe}]$ rises tightly with \FeH\ with no
        offset among P1, Im, and P2, and [N/Fe] correlates tightly with age across all three
        populations. However, [Na/Fe], [O/Fe], and [Mg/Fe] follow a separate track for each
        population, and P2 is offset to higher [Ba/Fe] at fixed \FeH\ \citep{Wang2026a}. P1 and P2
        overlap widely in \FeH, and the large helium enhancement is already present in the most
        metal-poor P2 stars \citep{Clontz2025a}.

      \item \textit{Mass budget.} Im and P2 together make up about two-thirds of the evolved stars
        within $r_h$ \citep{Clontz2026}.

      \item \textit{Debris.} In the halo substructures that \citet{Souza2026a} associate with the
        host of \ocen, P1-like stars dominate the outermost, earliest-stripped debris and follow the
        metallicity-spread sequence of dwarf galaxies, while the He--N--Al-rich P2-like stars are
        concentrated toward the inner regions and absent from the outskirts.

\end{itemize}}

  The kinematics can test whether the enriched stars formed dissipationally, from gas that cooled
  and settled in the cluster potential, or were assembled dissipationlessly from stars that had
  already formed. Stars that form from gas settled in a rotating potential inherit its angular
  momentum, and in simulations of second-generation formation the enriched stars form a faster
  rotating, flattened component \citep{Bekki2010, Henault-Brunet2015, Lacchin2022}. In M54 and the
  Milky Way nuclear star cluster, the metal-rich stars likewise have more rotational support than
  the metal-poor stars, which are possibly accreted \citep{Alfaro-Cuello2020, Kacharov2022b,
  Feldmeier-Krause2025b}. Stars assembled dissipationlessly, like these accreted components, are
  instead on radially biased orbits with less angular momentum \citep{Alfaro-Cuello2020,
  Kacharov2022b}. The kinematics therefore argue against P2 and Im forming in a long-lived, cold
  disk, and favor either in-situ formation in which the disk is disrupted early
  \citep[e.g.,][]{Lacchin2022}, or assembly from clumps that arrived as stars.
  Figure~\ref{fig:formation-schematic} shows these two commonly proposed scenarios.

  \subsubsection{Scenario 1: Hierarchical Assembly}
  Under hierarchical assembly, P2/Im form as separate stellar clumps or star clusters that spiral to
  the center of the dwarf via dynamical friction (Figure~\ref{fig:formation-schematic}, top row; see
  also \citealt{Fahrion2021, Fahrion2022a}), as in models that build nuclear star clusters from
  inspiraling GCs \citep{Antonini2012, Tsatsi2017, Neumayer2020}. Hierarchical assembly can take
  several forms:  Im and P2 both form in clumps detached from P1 and then fall in; only Im arrives
  as inspiraling GCs, and P2 forms afterward from the gas those clusters deliver to the center
  \citep{Mason2026}; sub-clusters merge while the cluster is still forming
  \citep[e.g.,][]{Lahen2025a}; or  a pre-formed, dense P2 system absorbs an infalling P1.
  
  Separate infalling clumps would also deliver some primordial stars. These would be counted as P1
  on the chromosome map, but would share the clump's age and orbits, whereas P1 is older than P2 at
  every \FeH\  and kinematically distinct from P2/Im. Im falls between P1 and P2 in age, an ordering
  that clumps forming and spiraling in one after another could reproduce, though only by
  coincidence. Individual clumps that condense from a larger chemically homogeneous parent cloud
  would inherit its composition, so the shared rise of $[(\mathrm{C+N+O})/\mathrm{Fe}]$
  \citep{Marino2012, Wang2026a} and the similar rise of the s-process elements ([La/Fe] in
  \citealt{Marino2012}, [Ba/Fe] in \citealt{Wang2026a}) with \FeH\ can be explained by a common
  birth environment. One kinematic feature may point toward separate assembly for Im: its rotation
  axis lies $\approx20^\circ$ from that of P2. In $N$-body models of nuclear star cluster growth by
  cluster inspiral, a late infall changes the direction of the cluster's angular momentum by
  $15$--$40^\circ$ \citep{Tsatsi2017}, so a population delivered late could retain a distinct
  rotation axis.
  
  If P1 instead fell onto a pre-formed, dense P2 system, the timeline is inverted. Because P2 is
  younger than P1 at every \FeH, P1 would then have formed elsewhere and fallen in only after P2 had
  formed. P1's flattening would then be acquired from the orbit of the infall instead of inherited
  from its birth environment: in the inspiral models of \citet{Tsatsi2017}, about two-thirds of the
  orbital angular momentum of the infalling clusters ends up in the nuclear cluster, which rotates
  and is flattened, although their models build the nuclear cluster from successive inspirals rather
  than a single infall. An infall that deposits enough orbital angular momentum to flatten P1 should
  also make it rotate faster, whereas P1 rotates more slowly than P2/Im at $200''$. Im has no
  natural place in this version, however, and the chemical continuity across P1, Im, and P2 would
  have to arise in separate systems.
  
  \subsubsection{Scenario 2: In-situ Gas Processing}
  Under in-situ gas processing (Figure~\ref{fig:formation-schematic}, bottom row), P2/Im instead
  form from gas retained within P1's own potential, without ever existing as a separate system.
  Because every population then draws on the same enrichment sequence, this scenario reproduces the
  shared $[(\mathrm{C+N+O})/\mathrm{Fe}]$--\FeH\ and [N/Fe]--age relations  without fine-tuning. A
  single enrichment sequence is harder to reconcile with the separate Na, O, and Mg tracks of each
  population, with the large overlap in \FeH\ between P1 and P2, which requires strong dilution of
  the enriched gas, and with the large helium enhancement already present in the most metal-poor P2
  stars. Variants in which Im forms by dilution \citep[e.g.,][]{Dondoglio2026} are also in tension
  with the ages, since Im is older than P2.
  
  The specific process that compresses and consumes the retained gas does not need to be pinned
  down. Compressive tides, as in galaxy mergers \citep{Renaud2009, Renaud2014}, tidally driven gas
  inflow to the nucleus \citep{Bekki2003}, or a purely internal cooling flow \citep{DErcole2008} all
  act primarily on gas rather than stars, and should leave the flattened dispersion field that P1
  inherited untouched while forming P2/Im at the cluster center.
  
  Gas that concentrates inside a rotating potential should settle into a disk \citep{Bekki2010}, and
  $N$-body models of second-generation disks in \ocen-like clusters with up to $\sim$30\% of the
  cluster mass become nearly spherical within the $\sim$12~Gyr of the simulations, while more
  massive disks remain flattened \citep{Mastrobuono-Battisti2013, Mastrobuono-Battisti2016}. The
  enriched populations together make up about two-thirds of the evolved stars within $r_h$, so a
  second-generation disk of this size would be expected to survive to the present day. A round P2/Im
  therefore suggests either that its gas was dynamically hot when it formed stars, or that it formed
  at densities high enough to erase a disk quickly, as when infalling pristine gas disrupts the disk
  in the high-density models of \citet{Lacchin2022}.

  \subsubsection{Debris, Mass Budget, and Timing}
  
  The debris of the disrupted progenitor argues against a late infall of P1 onto P2. P1 has
  counterparts across the host galaxy, whereas the enriched stars are found only near its center
  \citep{Souza2026a}. This favors a P1 that formed as part of the host's own stellar population,
  with its flattening inherited from the host's inner, rotating body, over a P1 delivered by a
  single infalling cluster. Finding the enriched stars only near the center of the halo debris does
  not by itself show that they formed in the nucleus. Clusters with multiple populations also form
  away from the nucleus: NGC~5634, associated with Sagittarius, shows the Na--O and Mg--Al
  anticorrelations typical of GCs \citep{Carretta2017}, and clusters that sink to the center by
  dynamical friction would deliver their enriched stars there. Both scenarios can therefore place
  the enriched stars at the center, provided that in scenario~1 the clumps arrived before losing
  many of their enriched stars to the outer debris. Neither scenario is excluded, but neither
  accounts for the chemistry, the ages, and the mass budget together without further assumptions.
  
  All of these scenarios share one problem, the mass budget: for a standard initial mass function,
  the polluters of a first generation cannot produce enough processed material to form enriched
  populations as numerous as those in \ocen\ \citep{Renzini2015, Bastian2018}. We do not attempt to
  solve this problem here, but note that a galactic nucleus, as \ocen\ likely was, eases it. Its
  deep potential retains the polluters' ejecta \citep{Souza2026a} and draws in fresh gas from the
  host \citep{Bekki2003}. Ejecta that collect at the center, where P2 now lies, are retained most
  easily. The enriched gas can also be gathered from the whole central region of the progenitor,
  whose outer layers were later stripped \citep{Souza2026a}. Its high stellar densities favor
  runaway stellar collisions \citep{PortegiesZwart2004a, Neumayer2020}, which can build supermassive
  stars \citep{Denissenkov2014} that process a large fraction of the cluster mass
  \citep{Gieles2018a}. Extremely massive stars that grow by accreting gas offer an alternative
  polluter \citep{Gieles2025a}. Runaway collisions can leave an IMBH \citep{PortegiesZwart2004a,
  Gieles2018a}, which could have grown into the one at the center of \ocen\ today
  \citep{Haberle2024}.
  
  Whichever scenario applies, the ages of the populations tie the enrichment to the infall of the
  progenitor. \citet{Laporte2026} date the pericentric passage of the GSE, the merger proposed to
  have brought \ocen\ into the MW, to $11.2\pm0.1$~Gyr from a coeval burst of in-situ globular
  cluster formation, and the GSE's most metal-rich clusters to $10.9\pm0.1$~Gyr. The most metal-rich
  stars of \ocen\ formed at $\approx$11~Gyr \citep{Clontz2024b}, so if \ocen\ was the GSE's nucleus,
  its enriched populations may have formed during the GSE's infall, and its star formation may have
  ceased once the GSE was disrupted \citep{Clontz2024b, Laporte2026, Mannu2026a}. In the simulations
  of \citet{Laporte2026}, the infalling satellite's central gas is compressed and forms stars on its
  final approach, before pericenter, which would supply an external trigger for the later bursts of
  enriched star formation.

  \begin{figure*}
    \centering
    \includegraphics[width=\linewidth]{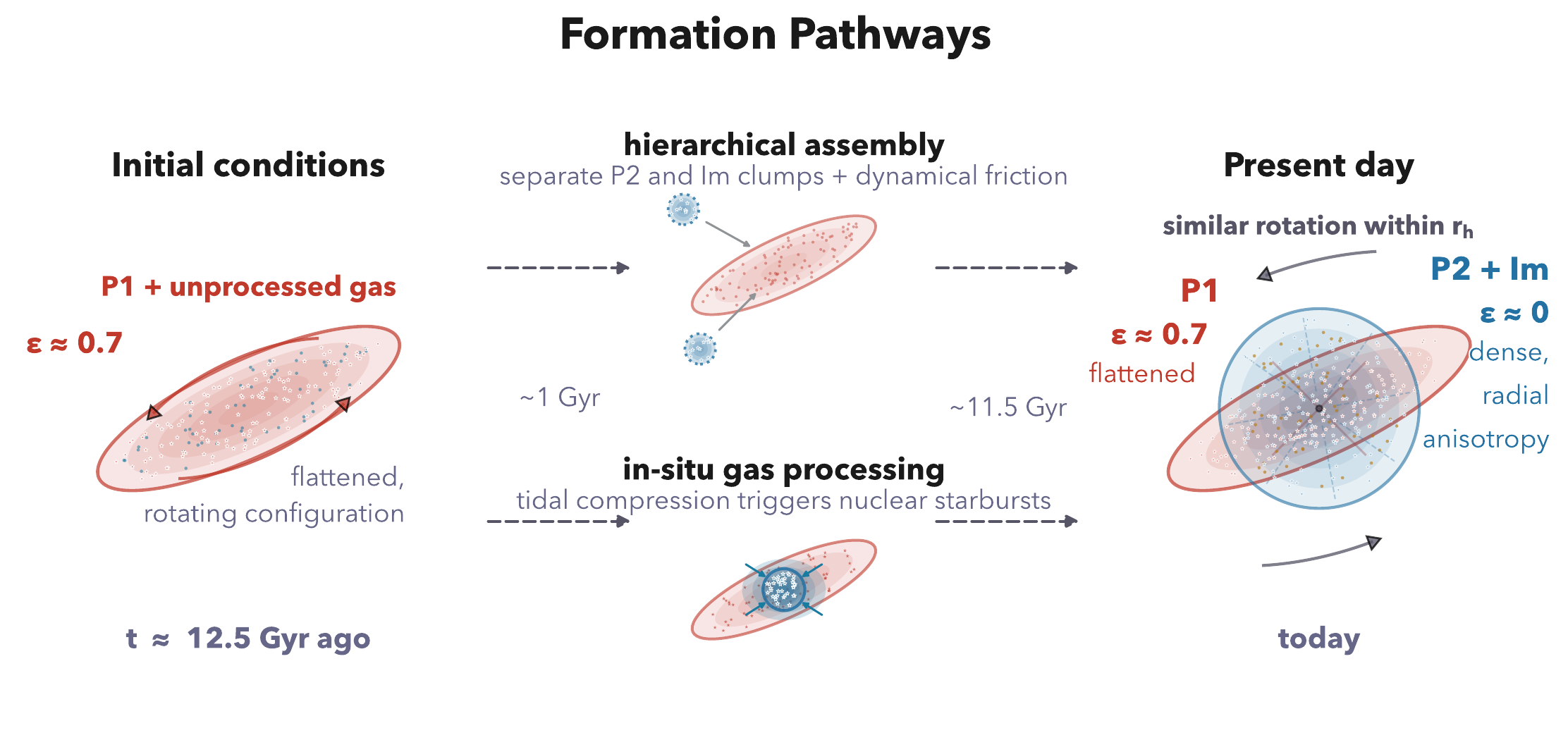}
    \caption{Schematic of the two formation scenarios for P2/Im discussed in this section, from P1's
      initial flattened, rotating configuration (\textit{left}) to the present day (\textit{right}).
      \textit{Top row:} under hierarchical assembly, P2/Im form as separate stellar clumps that
      spiral into P1's potential via dynamical friction. \textit{Bottom row:} under in-situ gas
      processing, P2/Im form directly from gas retained within P1's own potential, without ever
      existing as a separate stellar system. Both pathways converge on the same present-day
      configuration (\textit{right}): a primordial population with a flattened dispersion field and
      a centrally concentrated, spherical, radially anisotropic enriched population with similar
      rotation within $r_h$. Both scenarios reproduce the kinematic constraints established in
      Section~\ref{sec:dichotomy_compiled}, although in-situ gas processing does so only if the disk
    formed by the dissipating gas is erased early.}
    \label{fig:formation-schematic}
  \end{figure*}

  \subsection{One enrichment event or two?}
  \label{sec:im_p2}
  The kinematics do not clearly separate Im from P2, but P2 is more centrally concentrated than Im,
  which is distributed like P1. Gas that was retained and settled at the center would form stars at
  least as concentrated as P2, so Im formed over a more extended region, before the enriched gas had
  settled as far inward, or was delivered from outside, as in the inspiraling-cluster model of
  \citet{Mason2026}. The chemistry also separates them: the separate Na, O, and Mg tracks and the
  age ordering indicate that P1, Im, and P2 followed distinct enrichment tracks, whether or not they
  drew on a common reservoir, and that the continuity is limited to C+N+O and the N--age relation.
  Distinct tracks need not mean distinct gas sources, however. A single accretion event can bring
  the same gas into compressive tidal regions at successive pericenters, as in galaxy mergers
  \citep{Renaud2009, Renaud2014}, and can drive repeated nuclear starbursts \citep{Bekki2003}.
  Kinematically, the data require no more than two components; the enriched component could have
  formed in one continuous episode or in successive bursts.

  \subsection{Limitations}
  \label{sec:limitations}

  There are a few limitations to the interpretation presented above.

  First, the projected kinematics constrain the shape of the P1 dispersion field but not its
  intrinsic geometry. The axial ratio, vertical thickness, and inclination-corrected rotation
  amplitude remain degenerate in projection, and recovering them requires dedicated
  population-resolved dynamical modeling.

  Second, the present-day central concentration and radial anisotropy of P2 reflect the outcome of
  P2's formation, namely that it formed centrally concentrated, but not the mechanism for how it
  formed. The cluster's dynamical age is itself uncertain, because \ocen\ may have expanded once its
  tidal field weakened (Section~\ref{sec:relaxation}), and its orbit may have been reshaped by
  resonances with the Galactic bar~\citep{Dillamore2026a, Mannu2026a}, which may also strip stars
  along the orbit \citep{Mannu2026a}. Both complicate the interpretation of P2's present-day
  structure, although such external perturbations act mainly on the cluster's outskirts and are
  unlikely to affect the kinematics we measure within $r_h$.

  Third, our population-resolved sample cannot determine which population hosts the counter-rotating
  component previously detected in the core of \ocen \citep{Pechetti2024a, Haberle2025}. The
  counter-rotation signal comes from the central $\sim$10--20$''$, covered mainly by
  MUSE's Narrow Field Mode, where most stars lie below the magnitude limit of our population
  assignment (Section~\ref{sec:pop_defs}).

  Fourth, our sample is limited to evolved stars (RGB and SGB), for which MUSE provides LOS
  velocities and metallicities. Extending the population decomposition to the lower main sequence,
  where the chromosome map pseudo-colors are available for all HST-detected stars without requiring
  MUSE spectra, would increase the number of kinematic tracers by an order of magnitude and allow
  tests of differential energy equipartition between populations, a step that the new JWST imaging
  of the cluster (M.~H\"aberle et al., in prep.) will make possible.

  \section{Conclusions}
  \label{sec:conclusions}

  We combined high-precision HST proper motions with MUSE line-of-sight velocities from the oMEGACat
  survey to analyze the 3D phase-space kinematics of the multiple stellar populations in \ocen. By
  modeling the 2D dispersion, rotation, and proper-motion anisotropy fields, we reach the following
  conclusions about the cluster's formation and assembly history:
  \begin{enumerate}

    \item \textbf{Star-by-star abundance variations trace the dynamically distinct components.} The
    chromosome-map division reveals a flattened P1 component ($\epsilon\approx0.7$ in both HST
    proper motions and MUSE LOS velocities) and a P2/Im component that is consistent with
    $\epsilon=0$ in the LOS, and not significantly flattened in the proper motions. Divisions by
    \FeH\ mix stars from both components, which averages out these signatures, while the AMR sample
    is confined to the core, where the dispersion field is too flat to constrain any flattening.
    \FeH\ and AMR position likely reflect the chemical evolution of the host galaxy
    \citep{Laporte2026}, while the chromosome map reflects enrichment within the cluster.

    \item \textbf{P1 has a distinct, flattened orbital structure.} Within $r_h$, P1 has a flattened
      dispersion field in both the LOS and the proper motions, consistent with orbits more strongly
      aligned with the rotation axis, and rotates more slowly than P2/Im at $200''$. At larger
      radii, P1 has a larger angular-momentum dispersion \citep{Ziliotto2026}, is less radially
      anisotropic \citep{Bellini2018, Ziliotto2026, Pagnini2026a}, and rotates faster at
      $\sim3.5\,r_h$ \citep{Bellini2018}. The flattening suggests that P1 formed in a rotating,
      flattened environment. Its faster outer rotation may reflect more angular momentum at birth,
      or the outward diffusion of the initially compact enriched stars \citep{Tiongco2019}.

    \item \textbf{P2 formed more centrally concentrated than P1.} P2/Im have dispersion fields
    consistent with spherical symmetry and are radially anisotropic beyond $\sim100''$. P2 is more
    centrally concentrated than P1, while Im is distributed like P1. The central concentration of
    P2 suggests that P2 formed in a more compact configuration than P1 in each of the scenarios of
      Section~\ref{sec:hierarchical_assembly}. The same compact configuration can also help to
      explain P2's chemical enrichment: a deeper potential well retains a larger fraction of
      enriched ejecta.
  \end{enumerate}

  Three extensions follow from these results. First, while the kinematic ellipticity ($\epsilon
  \approx 0.7$) reported here is consistent with P1 forming a flattened structure, the precise 3D
  geometry (the intrinsic axial ratio, vertical thickness, and inclination-corrected rotation
  amplitude) remains unconstrained. Dedicated population-dynamical modeling of the 3D kinematics
  will be needed to recover the geometry of the P1 structure within the broader cluster. We will
  pursue this with population-resolved chemodynamical models in an upcoming paper (P. Smith et al.,
  in prep.).

  Second, these differences provide boundary conditions for formation modeling. Population-resolved
  $N$-body simulations that track all three populations self-consistently over a Hubble time could
  constrain the cluster's initial conditions and formation history. A forward-modeling approach that
  reproduces the present-day kinematics could constrain the formation of this nucleus and, by
  extension, the assembly history of its progenitor dwarf galaxy.

  Finally, the current study is limited to evolved stars (RGB and SGB) where kinematic tracers are
  luminous enough for reliable PM and LOS measurements. Extending the sample to fainter MS stars
  that lack reliable MUSE spectra would increase the number of stars by an order of magnitude and
  allow us to search for effects like differential energy equipartition between the populations. HST
  and JWST pseudo-colors can already separate the populations on the MS \citep{Bellini2017a,
  Scalco2025b, Ziliotto2026}. New JWST imaging of the cluster (M.~H\"aberle et al., in prep.) will
  allow us to extend this approach to the lower MS in a future paper.

  \begin{acknowledgments}
    P.J.S. is a fellow of, and acknowledges financial support from, the International Max Planck
    Research School for Astronomy and Cosmic Physics at the University of Heidelberg (IMPRS-HD).

    It is a pleasure to thank the MPIA Galactic Nuclei and Milky Way groups for helpful discussions
    and feedback on this work. This work further benefited from discussions with the participants of
    the Galactic Nuclei Astrophysical Phenomena workshop at the Sexten Center for Astrophysics.

    We thank Emanuele Dondoglio for kindly sharing with us the radial population fractions they
    measure in \citet{Dondoglio2026}, which we plot in Figure~\ref{fig:radial-fractions-chromo}.

    A.F.K. acknowledges funding from the Austrian Science Fund (FWF) [grant DOI 10.55776/ESP542].

    This research has made use of the Astrophysics Data System, funded by NASA under Cooperative
    Agreement 80NSSC21M00561. This work makes use of the colormaps in the CMasher package
    \citep{2020JOSS....5.2004V, CMasher_14186007}. Software citation information was aggregated
    using \href{https://flatironinstitute.github.io/software-citation-station/}{The Software
    Citation Station} \citep{software-citation-station-paper, software-citation-station-zenodo}.

    Anthropic's Claude models were used to assist with code development, literature searches, and
    language editing of the manuscript. All generated content was reviewed and edited by the
    authors.

  \end{acknowledgments}

  \facilities{VLT:Yepun(MUSE), HST(ACS, WFC3)}

  \software{\texttt{astropy} \citep{TheAstropyCollaboration2013,
    TheAstropyCollaboration2018, TheAstropyCollaboration2022, astropy_21262391}, \texttt{ArviZ}
    \citep{arviz_2019, ArviZ_21885631}, \texttt{corner} \citep{Foreman-Mackey2016}, \texttt{JAX}
    \citep{jax2018github}, \texttt{jax-cosmo} \citep{Campagne2023}, \texttt{matplotlib}
    \citep{Hunter2007}, \texttt{numpy} \citep{Harris2020}, \texttt{numpyro} \citep{Phan2019,
    bingham2019pyro}, \texttt{pandas} \citep{mckinney-proc-scipy-2010, pandas_23955360},
  \texttt{python} \citep{python}, and \texttt{scipy} \citep{Virtanen2020, scipy_22051554}}

  \appendix
  \counterwithin*{figure}{section}\counterwithin*{table}{section}
  \renewcommand{\thefigure}{\thesection\arabic{figure}}\renewcommand{\thetable}{\thesection\arabic{table}}
  \renewcommand{\theHfigure}{\thesection\arabic{figure}}\renewcommand{\theHtable}{\thesection\arabic{table}}

  \section{Sample Size per Subpopulation}\label{app:cuts}

  Table~\ref{tab:cuts} lists the number of stars in each subpopulation after each quality cut of
  Section~\ref{sec:kinquality}. The first row of each block counts the stars that meet the
  requirements of that division: valid chromosome-map indices and $12.9<m_{\rm F814W}<16.1$; a
  fiducial age and \FeH\ from \citet{Clontz2024b}; or $m_{\rm F625W}<17$ with a MUSE metallicity.
  The cuts are cumulative: the third row gives the stars that pass both, and ``Retained'' the number
  as a fraction of the first row. For the AMR, the last row gives the stars used in the fits. We
  note that no population is preferentially removed by the cuts.

  \begin{table}
    \centering
    \caption{Number of stars in each subpopulation after each quality cut. ``Retained'' gives the
      fraction of the first row that survives both cuts. The Metal-Rich AMR group is excluded from
    the fits (Section~\ref{sec:res_amr}).}
    \label{tab:cuts}
    \footnotesize
\begin{tabular}{lrrrrrr}
\hline
\multicolumn{7}{l}{\textit{Chromosome map}} \\
Cut & P2 & Im & P1 &  &  & Total \\
\hline
Population defined & 2323 & 1187 & 1531 &  &  & 5041 \\
Good astrometry \& membership & 2114 & 1079 & 1363 &  &  & 4556 \\
Good LOS velocity & 1904 & 940 & 1224 &  &  & 4068 \\
Retained & 82\% & 79\% & 80\% &  &  & 81\% \\
\hline
\multicolumn{7}{l}{\textit{AMR}} \\
Cut & Tight & Diffuse & Metal-Rich &  &  & Total \\
\hline
Population defined & 3448 & 3313 & 1008 &  &  & 7769 \\
Good astrometry \& membership & 3094 & 3024 & 847 &  &  & 6965 \\
Good LOS velocity & 2573 & 2496 & 653 &  &  & 5722 \\
Retained & 75\% & 75\% & 65\% &  &  & 74\% \\
Used in fits & 2573 & 2496 & -- &  &  & 5069 \\
\hline
\multicolumn{7}{l}{\textit{[Fe/H] quintiles}} \\
Cut & Met 1 & Met 2 & Met 3 & Met 4 & Met 5 & Total \\
\hline
Population defined & 1899 & 1730 & 1727 & 1709 & 1723 & 8788 \\
Good astrometry \& membership & 1575 & 1494 & 1464 & 1454 & 1411 & 7398 \\
Good LOS velocity & 1239 & 1237 & 1237 & 1238 & 1238 & 6189 \\
Retained & 65\% & 72\% & 72\% & 72\% & 72\% & 70\% \\
\hline
\end{tabular}

  \end{table}

  \section{Chromosome Map: 1D Kinematic Profiles}\label{app:chromo-1d}

  For the 1D profiles, we use both discrete, unbinned methods and binned profiles, which serve as a
  check on each other. The discrete method adopts the same Plummer dispersion and Lynden-Bell
  rotation models used for the 2D fields (Section~\ref{sec:modeling}), but uses only the radial
  position of each star rather than its full 2D position. For the binned profiles, we use adaptive
  logarithmic radial bins \citep[see, e.g.,][]{Kamann2018, Haberle2025}, requiring a minimum of 150
  stars per bin, and compute the dispersion and rotation within each bin, again implementing the
  likelihoods in JAX and sampling with NUTS. The binned profiles impose no functional form on the
  kinematics, allowing us to verify that the discrete models are appropriate for the data.

  In one dimension, the chromosome-map populations already hint at the P1/P2 kinematic dichotomy but
  are more difficult to interpret than the 2D maps. For P1, the LOS dispersion profile is lower in
  the core and flatter with radius than those of Im and P2, and the PM profiles show only modest
  differences between populations. These 1D trends are therefore best viewed as the radial
  projection of the stronger azimuthal structure seen in the 2D fields, rather than as independent
  evidence for a separate dynamical component. We compare each population with the other two
  populations combined over six radial bins. Across $\sigma_{\rm LOS}$, $V_{\rm rot}$,
  $\sigma_{\mu_R}$, $\sigma_{\mu_T}$, and the anisotropy, Im and P2 are consistent within
  uncertainties, except possibly in rotation axis; only P1 departs from the other populations, in
  $\sigma_{\rm LOS}$ and $V_{\rm rot}$. The $V_{\rm rot}$ offset is in the sense of slower P1
  rotation, as in the 2D fits (Section~\ref{sec:res_chromo}); its PM dispersions and anisotropy
  agree. The 1D profiles therefore do not distinguish the populations with the same clarity as the
  2D maps.

  \begin{figure*}
    \centering
    \includegraphics[width=\linewidth]{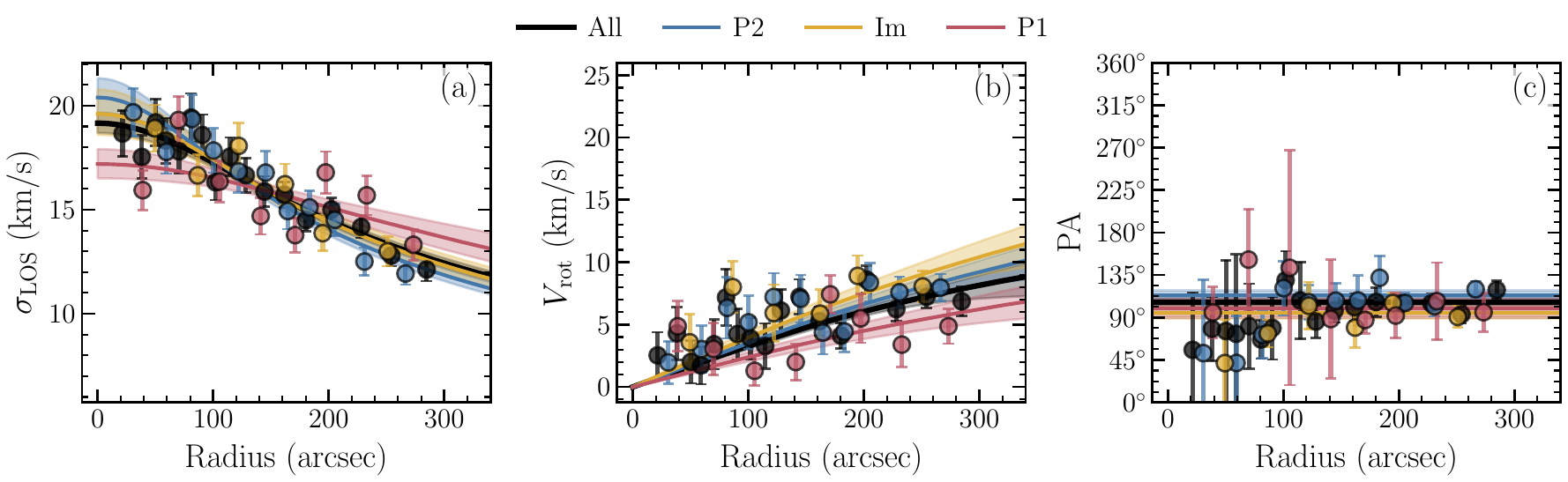}
    \caption{1D LOS kinematic profiles for the chromosome-map populations P1 (red), Im (yellow), and
      P2 (blue) and the bulk sample (black), as a function of projected radius. (a) LOS velocity
      dispersion $\sigma_{\rm LOS}(R)$, with best-fit Plummer model and 1$\sigma$ posterior bands.
      (b) LOS rotation velocity $V_{\rm rot}(R)$ from the Lynden-Bell model. (c) Position angle of
      the rotation. In all panels, we additionally show the binned version of the profiles for
      comparison. The P1 dispersion profile is lower in the core and flatter with radius than those
      of Im and P2, while the rotation axis of Im may be offset by $\approx20^\circ$ from that of P2
    (Section~\ref{sec:res_chromo}), and P1 rotates more slowly.}
    \label{fig:chromo-los}
  \end{figure*}

  \begin{figure*}
    \centering
    \includegraphics[width=\linewidth]{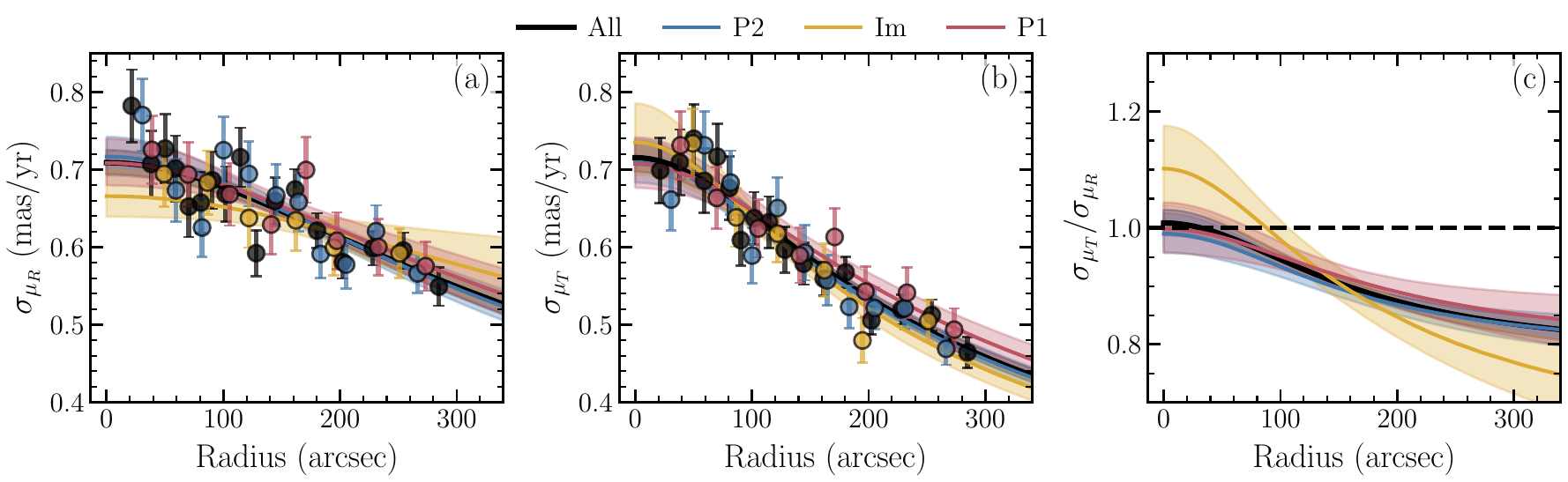}
    \caption{1D proper-motion kinematic profiles for the chromosome-map populations P1 (red), Im
      (yellow), and P2 (blue) as well as the bulk sample (black), as a function of projected radius.
      (a) Radial PM dispersion $\sigma_{\mu_R}(R)$. (b) Tangential PM dispersion
      $\sigma_{\mu_T}(R)$. (c) Anisotropy ratio $\sigma_{\mu_T}/\sigma_{\mu_R}$. All populations
    agree within the uncertainties.}
    \label{fig:chromo-pm}
  \end{figure*}

  \section{Metallicity Bin Kinematics}\label{app:metallicity-bins}

  The 1D kinematic profiles of the five \FeH\ quintiles show no significant differences in
  dispersion, rotation, or rotation-axis orientation. The 1D profiles are shown in
  Figures~\ref{fig:MH-los} and \ref{fig:MH-pm}. They are consistent with the 2D ellipticity fits
  (Figure~\ref{fig:MH-ellipticity-violin}), and indicate
  that the kinematics of the \FeH\ quintiles are consistent with each other within the
  uncertainties.
  \begin{figure*}
    \centering
    \includegraphics[width=\linewidth]{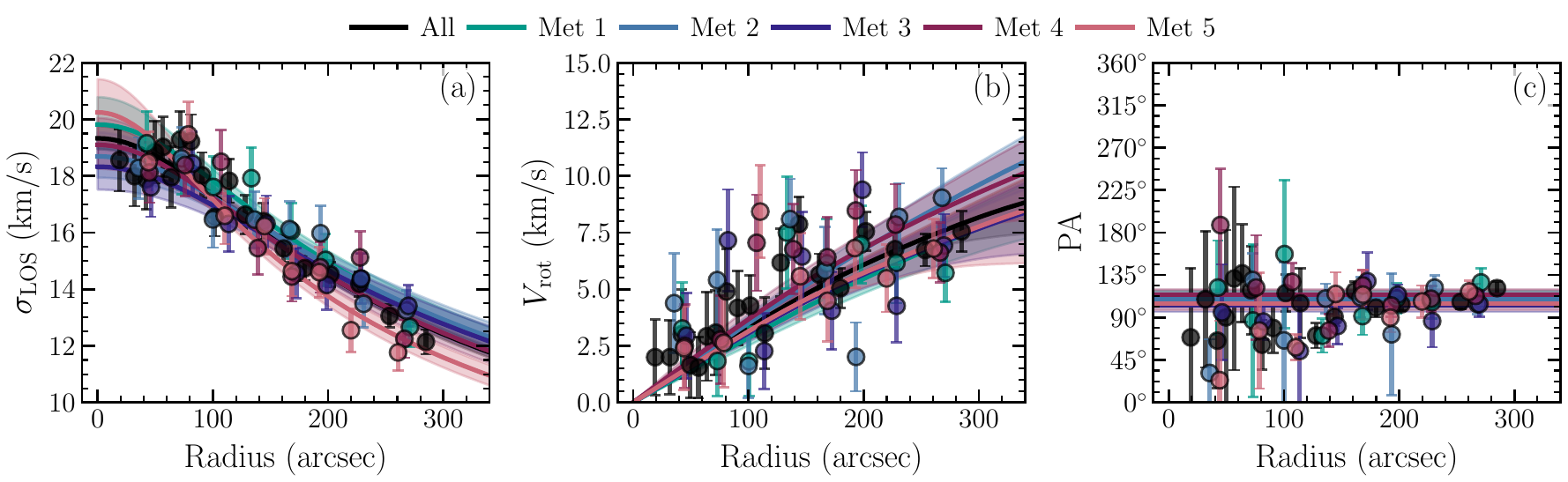}
    \caption{1D LOS kinematic profiles for the five \FeH\ quintile populations (colored
      from most metal-poor to most metal-rich) and the bulk sample (black). (a) LOS velocity
      dispersion $\sigma_{\rm LOS}(R)$. (b) LOS rotation velocity $V_{\rm rot}(R)$. (c) Position
      angle of the rotation. No significant differences in dispersion or rotation are detected
    across the quintiles.}
    \label{fig:MH-los}
  \end{figure*}

  \begin{figure*}
    \centering
    \includegraphics[width=\linewidth]{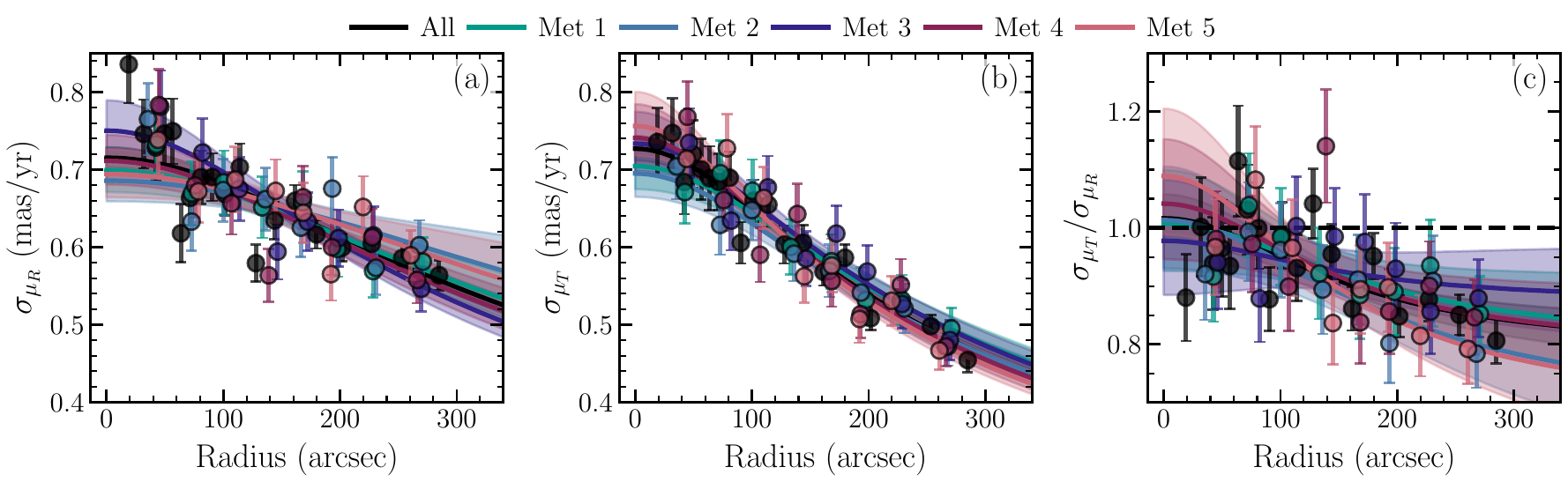}
    \caption{1D proper-motion kinematic profiles for the five \FeH\ quintile populations (same color
      scheme as Figure~\ref{fig:MH-los}). (a) Radial PM dispersion $\sigma_{\mu_R}(R)$. (b)
      Tangential PM dispersion $\sigma_{\mu_T}(R)$. (c) Anisotropy ratio
      $\sigma_{\mu_T}/\sigma_{\mu_R}$. Dispersion and anisotropy profiles are consistent across all
    metallicity quintiles within the uncertainties.}
    \label{fig:MH-pm}
  \end{figure*}

  \section{Age--Metallicity Relation Kinematics}\label{app:amr}

  Figures~\ref{fig:AMR-pm}, \ref{fig:AMR-los}, \ref{fig:AMR-pm-maps}, and \ref{fig:AMR-los-maps}
  show the 1D profiles and 2D maps of the Tight and Diffuse AMR populations. These maps are harder
  to interpret than the chromosome-map results because of their radial coverage: the AMR requires
  F606W photometry, which confines the sample to the cluster core (Section~\ref{sec:res_amr}).
  Across this range the dispersion varies only a little, and the few stars in the outermost bins
  dominate any measured gradient in the discrete fits. Within these limits, the Tight and Diffuse
  populations agree with each other and with the bulk in PM dispersion and anisotropy, and in LOS
  rotation amplitude and position angle. The one visible difference is the LOS dispersion of the
  Diffuse population, which stays nearly flat across the field while the Tight population follows
  the bulk decline (Figure~\ref{fig:AMR-los}). This difference rests on the sparsely populated outer
  bins and has no counterpart in the 2D ellipticity fits (Figure~\ref{fig:amr-ellipticity-violin}).

  \begin{figure*}
    \centering
    \includegraphics[width=\linewidth]{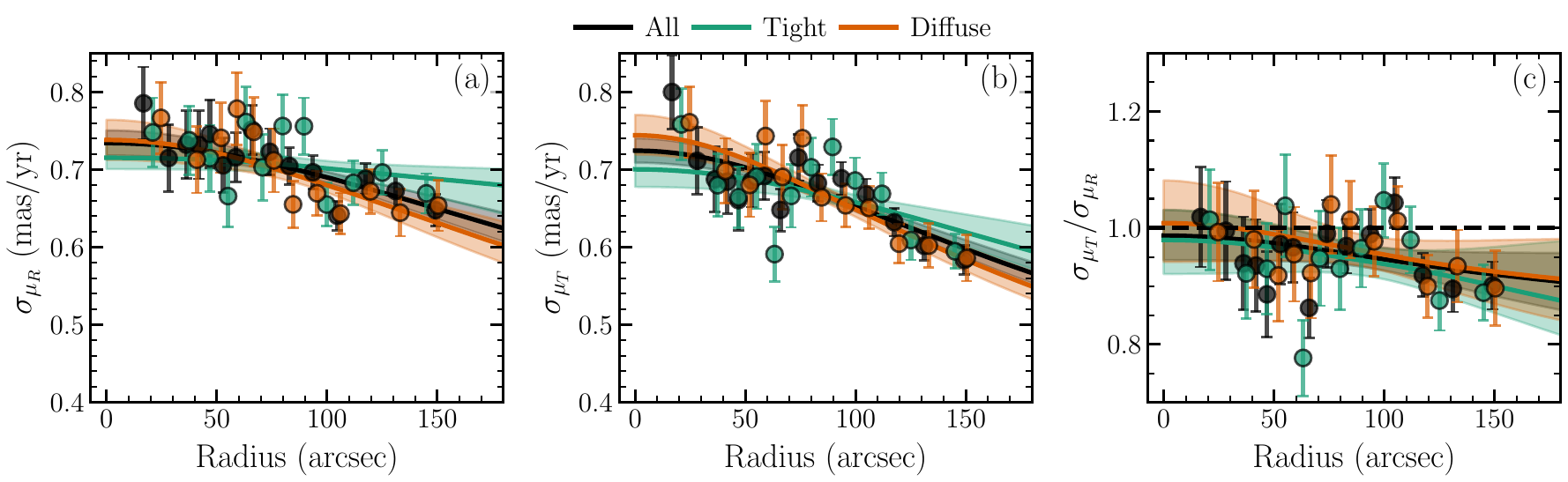}
    \caption{1D proper-motion kinematic profiles for the bulk AMR sample (black), Tight (green), and
      Diffuse (orange) populations. Panel layout as in Figure~\ref{fig:chromo-pm}. The two
    populations show broadly similar PM dispersion and anisotropy profiles.}
    \label{fig:AMR-pm}
  \end{figure*}

  \begin{figure*}
    \centering
    \includegraphics[width=\linewidth]{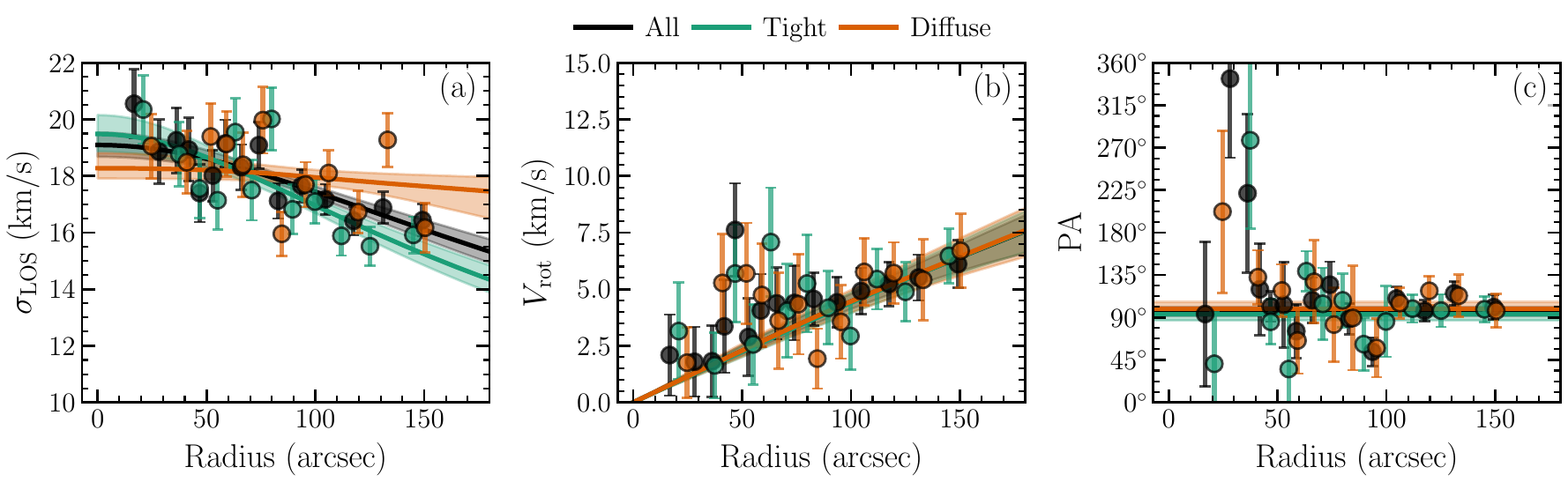}
    \caption{1D LOS kinematic profiles for the bulk AMR sample (black), Tight (green), and Diffuse
      (orange) populations. Panel layout as in Figure~\ref{fig:chromo-los}. The Diffuse population
      shows a largely flat dispersion profile across the radial range probed, while the Tight
      population more closely tracks the overall cluster dispersion decline. No significant
    differences in rotation amplitude or position angle are detected.}
    \label{fig:AMR-los}
  \end{figure*}

  \begin{figure*}
    \centering
    \includegraphics[width=\linewidth]{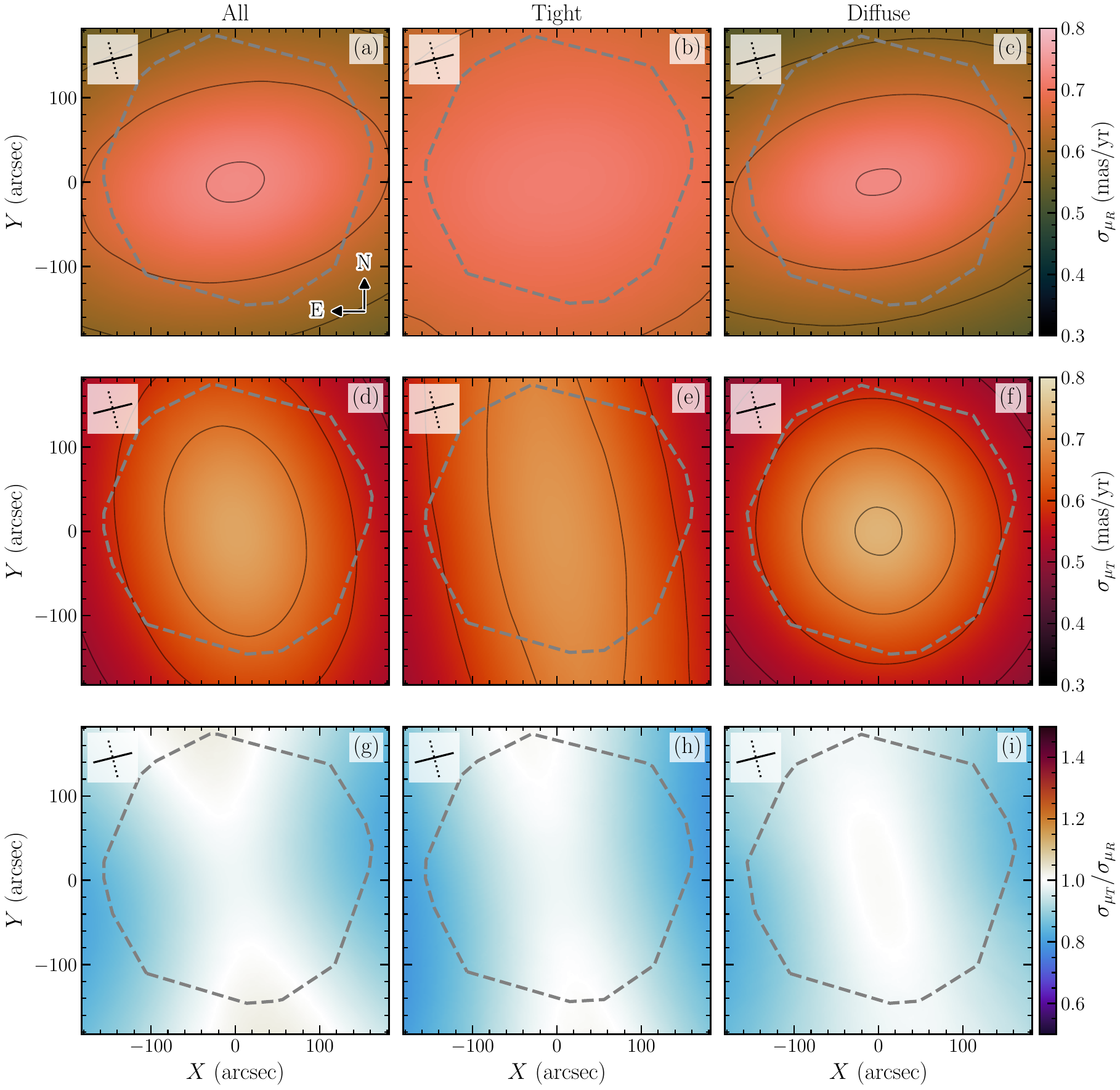}
    \caption{2D HST proper-motion kinematic maps for the AMR sample, Tight, and Diffuse populations
      (columns, left to right). Rows and axis lines as in Figure~\ref{fig:chromo-maps}. Because the
      maps cover only the cluster core, where the dispersions vary little across the field, the
      shapes of these fields are weakly constrained; this includes the elongated Tight
    $\sigma_{\mu_T}$ field (panel e).}
    \label{fig:AMR-pm-maps}
  \end{figure*}

  \begin{figure*}
    \centering
    \includegraphics[width=\linewidth]{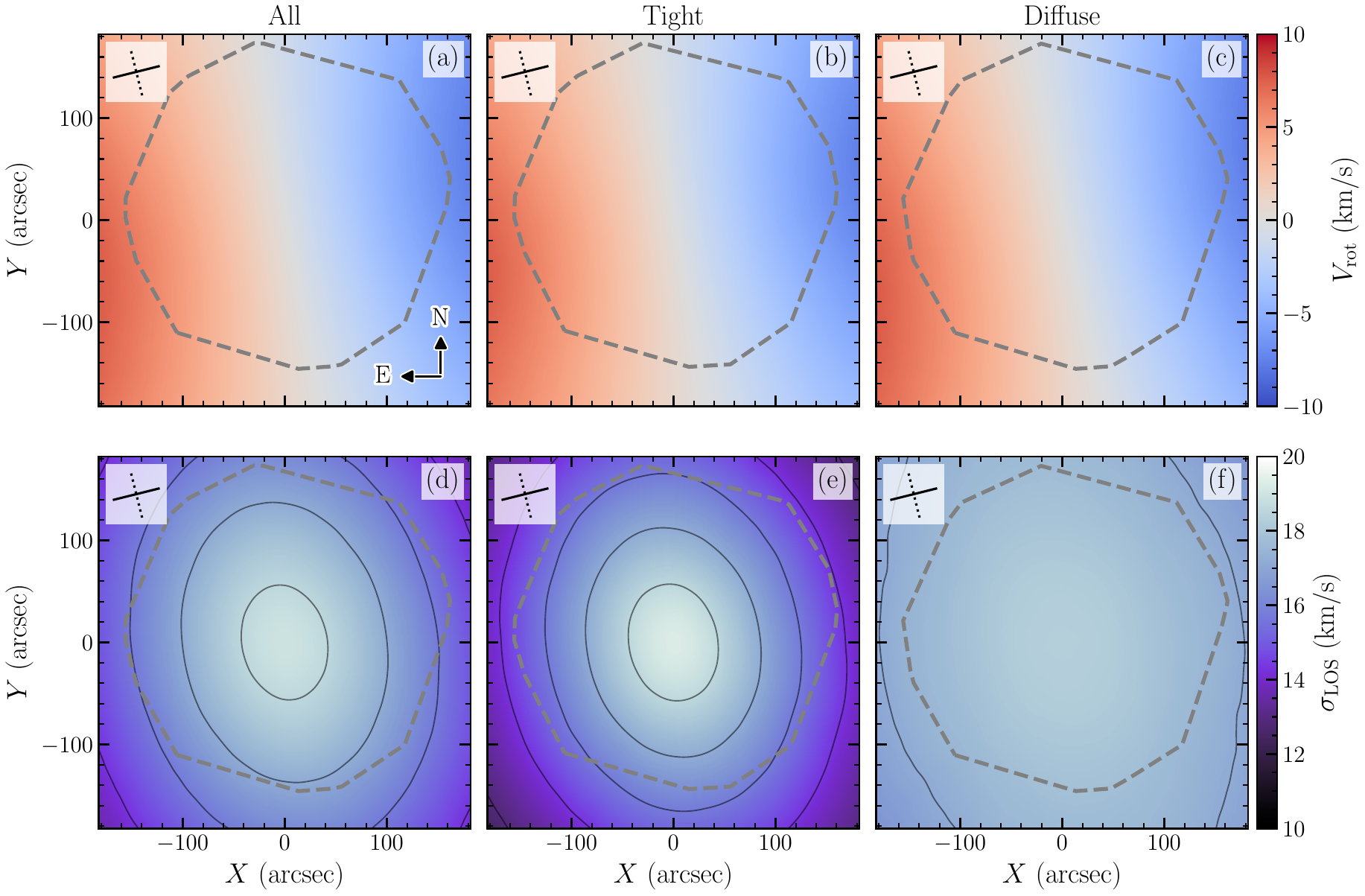}
    \caption{2D MUSE LOS rotation and velocity dispersion maps for the AMR sample, Tight, and
      Diffuse populations (columns, left to right). Rows and axis lines as in
      Figure~\ref{fig:chromo-los-maps}. No significant kinematic flattening is detected; ellipticity
    posteriors are consistent with $\epsilon=0$ in both populations.}
    \label{fig:AMR-los-maps}
  \end{figure*}

  Figure~\ref{fig:amr-ellipticity-violin} shows the dispersion-field ellipticity
  posteriors of the Tight and Diffuse AMR populations.

  \begin{figure*}
    \centering
    \includegraphics[width=\linewidth]{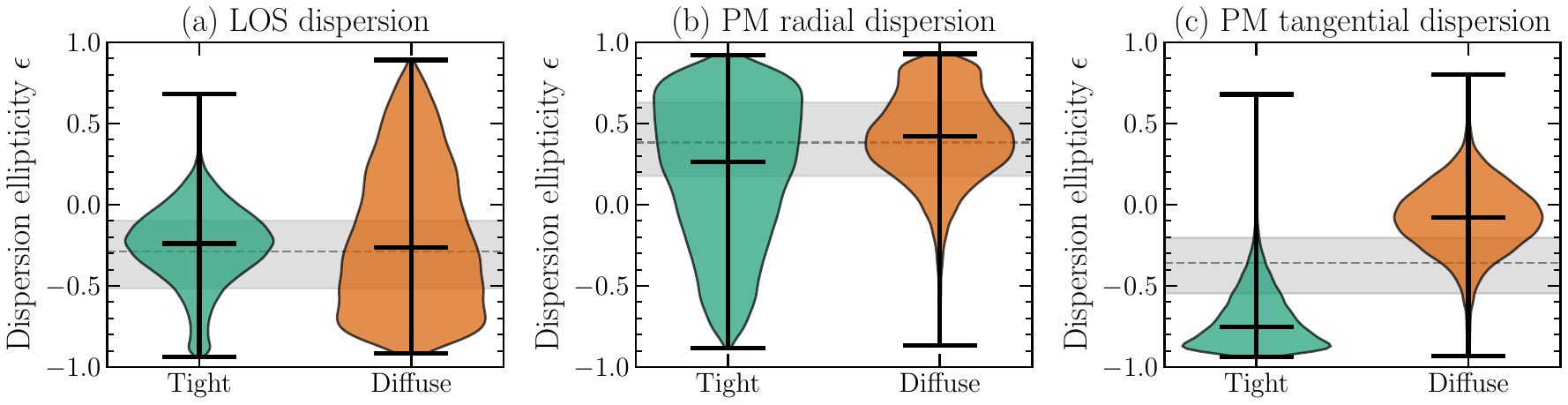}
    \caption{As in Figure~\ref{fig:chromo-ellipticity-violin}, but for the Tight and Diffuse AMR
      populations, compared against the bulk AMR sample posterior. Because the AMR sample lies
      almost entirely inside the core radius, where the dispersion field is nearly flat, every
      posterior, including that of the bulk, is broad and weakly constrained
    (Section~\ref{sec:res_amr}).}
    \label{fig:amr-ellipticity-violin}
  \end{figure*}

  \section{Rotation Posteriors}\label{app:rotation}

  Figure~\ref{fig:rotation-violin} collects the LOS rotation posteriors of all three divisions.
  Because the rotation curve peaks at the edge of our field, the projected peak amplitude $A_{\rm
  rot}$ is degenerate with $R_{\rm peak}$ and its posteriors are broad. The rotation velocity on the
  kinematic major axis at $R=200''$, $V_{\rm rot}(200'')$, lies inside the field and is computed for
  each posterior draw from the same fits, so it is better constrained; it is the rotation quantity
  we compare in the main text (Section~\ref{sec:modeling}).

  \begin{figure*}
    \centering
    \includegraphics[width=\linewidth]{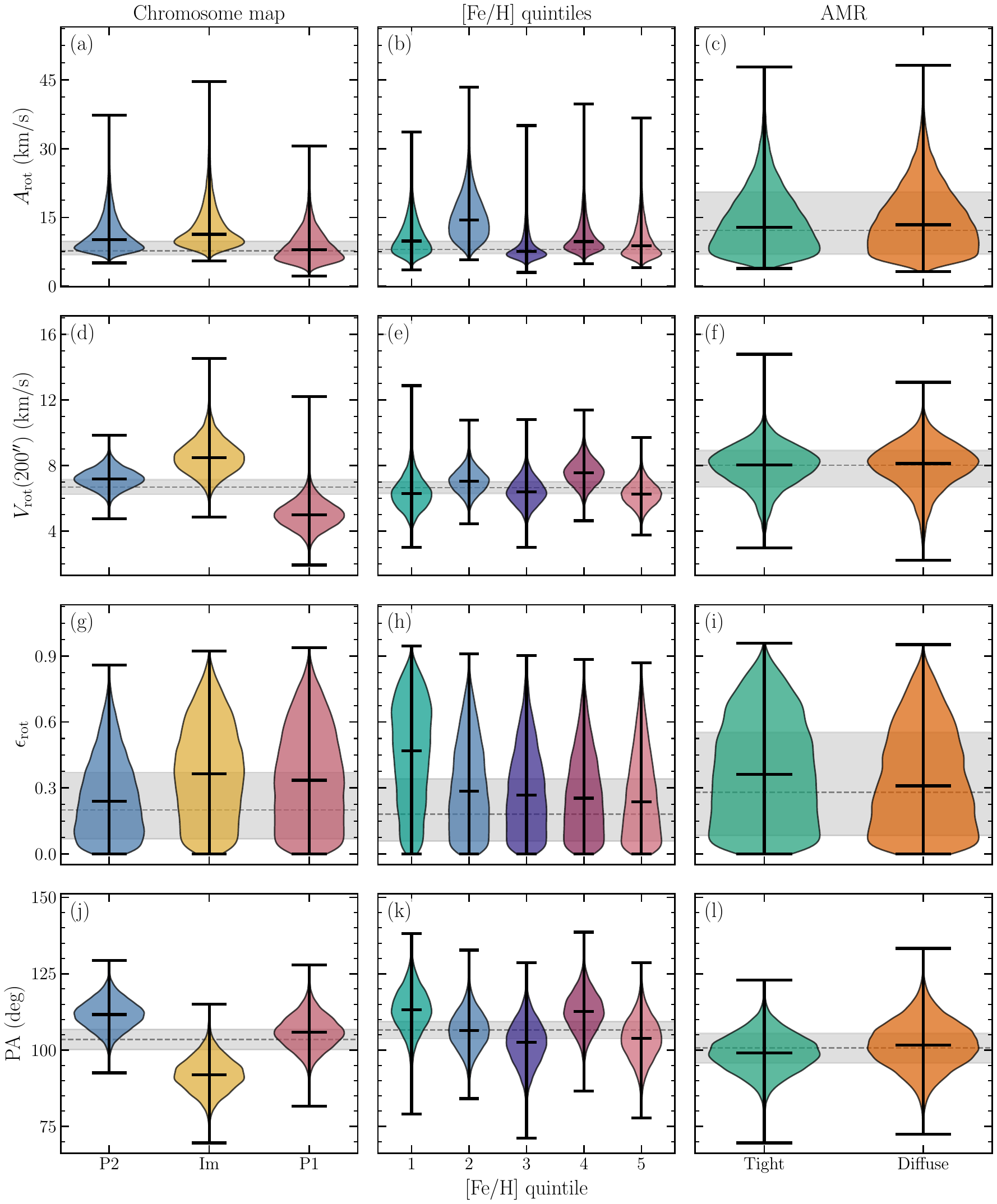}
    \caption{Posterior distributions of the LOS rotation parameters for the chromosome-map
      populations (left), the \FeH\ quintiles (middle), and the Tight and Diffuse AMR populations
      (right). \textit{Rows, top to bottom:} projected peak rotation amplitude $A_{\rm rot}$ (a--c),
      rotation velocity on the kinematic major axis at $R=200''$ derived from the same posterior
      draws (d--f), rotation-field ellipticity $\epsilon_{\rm rot}$ (g--i), and position angle of
      the kinematic major axis, east of north (j--l). Bulk posteriors and bars as in
    Figure~\ref{fig:chromo-ellipticity-violin}.}
    \label{fig:rotation-violin}
  \end{figure*}
  \bibliography{Research,software}
  \bibliographystyle{aasjournal}
\end{CJKwrap}
\end{document}